\documentclass[]{pasj01}

\usepackage{url}
\usepackage{mystyle}

\begin{document}
\title{A new \emph{off-point-less} observing method for millimeter and submillimeter
       spectroscopy with a frequency-modulating local oscillator (FMLO)}

\author{Akio \textsc{Taniguchi}\altaffilmark{1,2}}
\author{Yoichi \textsc{Tamura}\altaffilmark{1}}
\author{Kotaro \textsc{Kohno}\altaffilmark{2,3}}
\author{Shigeru \textsc{Takahashi}\altaffilmark{4}}
\author{Osamu \textsc{Horigome}\altaffilmark{5}}
\author{Jun \textsc{Maekawa}\altaffilmark{4}}
\author{Takeshi \textsc{Sakai}\altaffilmark{6}}
\author{Nario \textsc{Kuno}\altaffilmark{4,7,8}}
\author{Tetsuhiro \textsc{Minamidani}\altaffilmark{4,9}}

\email{taniguchi@a.phys.nagoya-u.ac.jp}

\altaffiltext{1}{Division of Particle and Astrophysical Science,
                 Graduate School of Science, Nagoya University,
                 Furocho, Chikusa-ku, Nagoya, Aichi 464-8602, Japan}
\altaffiltext{2}{Institute of Astronomy, The University of Tokyo,
                 2-21-1 Osawa, Mitaka, Tokyo 181-0015, Japan}
\altaffiltext{3}{Research Center for Early Universe,
                 School of Science, The University of Tokyo,
                 7-3-1 Hongo, Bunkyo-ku, Tokyo 113-0033, Japan}
\altaffiltext{4}{Nobeyama Radio Observatory,
                 National Astronomical Observatory of Japan (NAOJ),
                 National Institutes of Natural Sciences (NINS),
                 462-2 Nobeyama, Minamimaki, Minamisaku, Nagano 384-1305, Japan}
\altaffiltext{5}{Zero Co. Ltd., Nagano, Nagano, Japan}
\altaffiltext{6}{Graduate School of Informatics and Engineering,
                 The University of Electro-Communications,
                 1-5-1 Chofugaoka, Chofu, Tokyo 182-8585, Japan}
\altaffiltext{7}{Graduate School of Pure and Applied Sciences,
                 University of Tsukuba,
                 1-1-1 Ten-nodai, Tsukuba, Ibaraki 305-8577, Japan}
\altaffiltext{8}{Tomonaga Center for the History of the Universe,
                 1-1-1 Tenodai, Tsukuba, Ibaraki 305-8571, Japan}
\altaffiltext{9}{Department of Astronomical Science, School of Physical Science,
                 SOKENDAI (The Graduate University for Advanced Studies),
                 2-21-1 Osawa, Mitaka, Tokyo 181-8588, Japan}


\maketitle

\begin{abstract}
  We propose a new observing method for single-dish millimeter and submillimeter spectroscopy using a heterodyne receiver equipped with a frequency-modulating local oscillator (FMLO).
  Unlike conventional switching methods, which extract astronomical signals by subtracting the reference spectra of off-sources from those of on-sources, the FMLO method does not need to obtain any off-source spectra; rather, it estimates them from the on-source spectra themselves.
  The principle is a high dump-rate (10~Hz) spectroscopy with radio frequency modulation (FM) achieved by fast sweeping of a local oscillator (LO) of a heterodyne receiver:
  Because sky emission (i.e., off-source) fluctuates as $1/f$-type and is spectrally correlated, it can be estimated and subtracted from time-series spectra (a timestream) by principal component analysis.
  Meanwhile astronomical signals remain in the timestream since they are modulated to a higher time-frequency domain.
  The FMLO method therefore achieves (1) a remarkably high observation efficiency, (2) reduced spectral baseline wiggles, and (3) software-based sideband separation.
  We developed an FMLO system for the Nobeyama 45-m telescope and a data reduction procedure for it.
  Frequency modulation was realized by a tunable and programmable first local oscillator.
  With observations of Galactic sources, we demonstrate that the observation efficiency of the FMLO method is dramatically improved compared to conventional switching methods.
  Specifically, we find that the time to achieve the same noise level is reduced by a factor of 3.0 in single-pointed observations and by a factor of 1.2 in mapping observations.
  The FMLO method can be applied to observations of fainter ($\sim$mK) spectral lines and larger ($\sim$deg$^{2}$) mapping.
  It would offer much more efficient and baseline-stable observations compared to conventional switching methods.
\end{abstract}

\section{Introduction}
\label{s:introduction}

Improving the sensitivity of a single-dish radio telescope system is always an important issue in modern observational astronomy, especially in the era of the Atacama Large Millimeter/submillimeter Array (ALMA).
Detecting faint molecular line emission by single-dish blind redshift spectroscopic surveys is essential to studying distant submillimeter galaxies (SMGs; e.g., \cite{Blain2002}) with great help of large collecting area (e.g., \cite{Yun2015}).
Efficient single-dish mapping spectroscopy is also important to ALMA itself as ALMA uses four single dish antennas (Total Power Array of the Atacama Compact Array) in order to improve the fidelity of interferometric images.

There are many factors that limit the sensitivity of ground-based spectroscopic observations with single-dish radio telescopes.
The standard-deviation noise level of a spectrum, $\Delta S$, by a standard position-switching observation is expressed as
\begin{equation}
  \Delta S
  = \frac{\sqrt{2}\,
          k\msb{B}\,
          T\msb{sys}}
         {A\,
          \eta\msb{ap}\,
          \sqrt{N\msb{pix}\,
                \Delta\nu\,
                t\msb{total}\,
                \eta\msb{obs}}},
  \label{eqn:sensitivity}
\end{equation}
where $k\msb{B}$, $T\msb{sys}$, $A$, $\eta\msb{ap}$, $N\msb{pix}$, $\Delta\nu$, and $t\msb{total}$ are the Boltzmann constant, a system noise temperature, the collecting area of an antenna, aperture efficiency, the number of feeds, frequency width of a spectroscopic channel, total observation time including any overheads, respectively.
$\eta\msb{obs}$ is observation efficiency defined as a fraction of on-source time, $t\msb{on}$, over $t\msb{total}$:
\begin{equation}
  \eta\msb{obs} \equiv \frac{t\msb{on}}{t\msb{total}}.
  \label{eqn:observation-efficiency}
\end{equation}
Enormous efforts (requiring a considerable amount of resources) have been made to improve $T\msb{sys}$ and the effective collecting area, $A\,\eta\msb{ap}$, or to increase $N\msb{pix}$ (e.g., \cite{Minamidani2016, Schuster2004}).
On the other hand, although the parameters related to the observing methods, such as the factor of $\sqrt{2}$ and $\eta\msb{obs}$, have rooms for improving the sensitivity, they have not been fully explored yet.

The conventional position switching (PSW) and frequency switching (FSW) methods have been widely used in single-pointed\footnote{We hereafter use the term ``single-pointed'' when a telescope tracks a celestial coordinates (i.e., an observation of a point-like source). In contrast to single-pointed, the term ``mapping'' is used when a telescope scans a certain region.} spectroscopic observations in (sub-)millimeter astronomy (\cite{Wilson2012}).
These switching methods are necessary to estimate and correct for bandpass gains and sky levels based on a comparison of reference spectra with a major assumption that the condition of the telescope (i.e., bandpass and receiver noise temperature) and atmosphere (i.e., opacity) can be regarded as being constant in the time interval between on- and off-points\footnote{We hereafter define the term ``on-point'' as the celestial coordinates of an astronomical source and ``off-point'' as the ones without any sources.} or frequency shift from one to another.
In both methods, however, making a comparison (i.e., subtraction) with a reference spectrum is virtually equivalent to an addition of noise to the on-point spectrum, which is why the factor of $\sqrt{2}$ is multiplied to the right side of equation~\ref{eqn:sensitivity}.

Another issue is the spectral baseline fluctuation across emission-free channels:
The incident sky emission is generally time-variable and inhomogeneous at the (sub-)millimeter wavelength \citep{Lay2000}.
When the switching periods between on- and off-points (or frequency shift) are longer than the typical time-scale of sky variations, imbalance between two spectra can cause baseline fluctuations in the resulting spectra, because the conventional chopper wheel method does not deal with \emph{in-situ} estimation of bandpass gains and sky levels.

The resulting $\eta\msb{obs}$ offered by the PSW method is therefore not so high ($0.1 \lesssim \eta\msb{obs} < 0.5$) because of off-point measurements, telescope slewing time between on- and off-points, and some ``flagging'' of bad spectra due to baseline fluctuations.
As an improvement of the PSW method, a novel method that uses a smoothed off-point bandpass \citep{Yamaki2012} in order to reduce the noise added by the subtraction is a good compromise to offer $0.5 < \eta\msb{obs} < 1$.
As for observing an extended region, the on-the-fly (OTF) mapping method (\cite{Sawada2008}) is more efficient because it continuously drives an antenna to cover the region rapidly, and measurements of the off-point are only taken between scans.
These improvements, however, still require off-point measurements, and the degrading of $\eta\msb{obs}$ is still possible.

On the other hand, the $\eta\msb{obs}$ offered by the FSW method is higher ($\eta\msb{obs} > 0.5$).
The targets of the method, however, are limited to narrow spectral features such as Galactic quiescent sources because the line width must be narrower than the frequency shift.
To improve the FSW method, \citet{Heiles2007} has proposed obtaining spectra with more than two frequencies and then directly solving and correcting for IF-dependent bandpass gains by least-squares fitting (least-squares frequency switch; LSFS).
This approach assumes that the RF spectral shape should remain constant throughout an observation, thus, any spectral undulation due to non-linear response to variable atmospheric emission and receiver gain will result in systematic errors.

In contrast, a method achieving a high observation efficiency ($\eta\msb{obs} \approx 0.9-1$) that never needs off-point measurements has been developed; furthermore, it has been extensively employed in recent deep extragalactic surveys and cosmic microwave background (CMB) experiments on the basis of ground-based facilities using multi-pixel direct detector cameras.
The output of a ground-based telescope is always dominated by the atmospheric emission.
If a receiver has array detectors (e.g., a multibeam receiver or a spectrometer), the output time-series data (timestream, hereafter) from the detectors are mutually correlated, because the detectors see almost the same part of the troposphere ($\sim 1$~km above the ground).
Because these \emph{correlated noises} are known to behave as $1/f$-type noises and have large power at low frequencies ($\lesssim 10$~Hz) in the timestream, filtering out the correlate modes of the timestream that are common among multiple detector outputs with, for example, principal component analysis (PCA), can provide estimates of \emph{in-situ} and remove the awkward low frequency noises induced mainly by the atmosphere (\cite{Laurent2005,Scott2008}).
At the same time, it is also important to modulate the astronomical signals involved in the timestream into higher frequency domains so as not to filter out the astronomical signals of interest \citep{Kovacs2008}.
In the continuum deep surveys and CMB experiments, this modulation is achieved by quickly moving the telescope pointing across the sky.

Here, we introduce the concept of correlated noises and their removal into (sub-)millimeter spectroscopy and propose a new observing method for \emph{in-situ} estimation of bandpass gains and sky levels.
If one considers the one-to-one correspondence between multibeam (i.e.,  detector array camera) imaging observations and spectroscopic observations, which are given by
\begin{eqnarray}
  \mr{detectors\ of\ a\ camera} \quad&\rightarrow&\quad \mr{channels\ of\ a\ spectrometer} \nonumber\\
  \mr{moving\ the\ pointing} \quad&\rightarrow&\quad \mr{sweeping\ the\ frequency,} \nonumber
\end{eqnarray}
then, this noise removal technique can be applicable to spectroscopic observations \citep{Kovacs2008,Tamura2013}.
In other words, the removal of correlated noise without measurements of off-point spectra can be introduced to (sub-)millimeter spectroscopy, if we obtain a time-series spectra (a timestream) at an on-point with its observed frequency modulated at a dump rate of $\gtrsim10$~Hz in order to capture the $1/f$-like noise behavior of the sky and to estimate and remove it.
This new observing method with a totally different operation principle is a frequency ``modulation'' (FM) method; modulation of an observing frequency can be achieved with a heterodyne receiver by fast sweeping of a local oscillator (LO) frequency using a digital signal generator.
We therefore call our proposed method a \emph{frequency-modulating local oscillator} (FMLO) method.
As the FMLO method is independent of antenna movement of a telescope, it offers both single-pointed and mapping capabilities.
The advantages of the FMLO method are as follows: (1) high observation efficiency ($\eta\msb{obs} \gtrsim 0.9$) because of no off-point integration; (2) reduction in baseline ripples because of in-situ off-point estimation by PCA; (3) sideband separation in an offline data reduction; and (4) low cost implementation because existing instruments are likely to be available for the FMLO method.

In this paper, we report the principle, instrumentation, and observational demonstration of the FMLO method.
We introduce the principle of the FMLO method in section~\ref{s:principle} with mathematical expression of a timestream.
In section~\ref{s:instrumentation}, we describe the FMLO system of a telescope and its requirements, and the data reduction procedure after an FMLO observation.
We then demonstrate single-pointed and mapping observations of the FMLO method for Galactic bright sources with the FMLO system on the Nobeyama 45-m telescope in section~\ref{s:demonstration}.
We also verify how the resulting spectra obtained with the FMLO method are consistent with those obtained with the conventional method with a remarkable improvement of $\eta\msb{obs}$.
Finally, we discuss the advantages and limitations of the FMLO method in section~\ref{s:discussion}.

\section{Principle}
\label{s:principle}

\begin{figure*}[t]
  \centering
  \includegraphics[width=\linewidth]{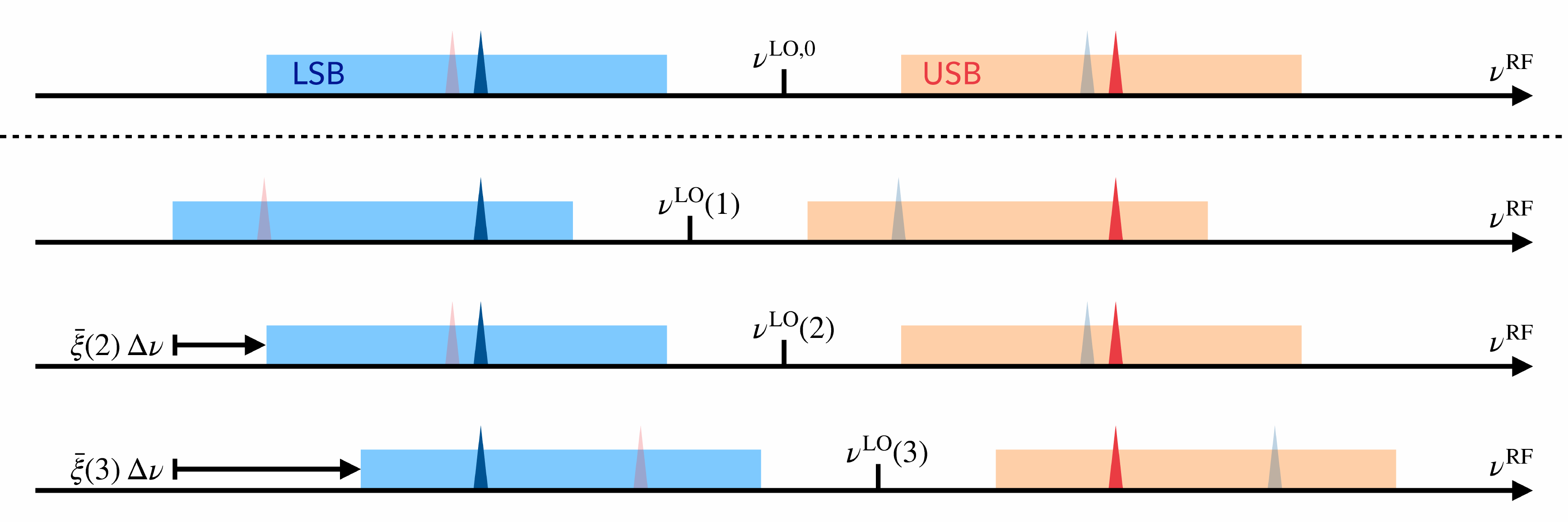}
  \caption{
    Schematic diagram showing modulation and demodulation of a timestream according to FM channels with mathematical expressions defined in section~\ref{subs:mathematical-expression-of-timestreams} and \ref{subs:modulation-and-demodulation-of-timestreams}.
    It supposes an FMLO observation of both USB and LSB (represented as colored boxes) where only one spectral line exists at each sideband (represented as colored spikes).
    Spikes with dimmed colors are the lines from the image sideband.
    A modulated timestream is an output of a spectrometer itself.
    On the other hand, a demodulated timestream is a matrix generated when we align each sample of the modulated timestream with the RF (sky) frequency (see also figure~\ref{fig:principle-2}).
    It is clear that a line in a sideband and that contaminated from the \emph{image} sideband are reversely frequency modulated, which enables us to achieve sideband separation in the post processing.
  }
  \label{fig:principle-1}
\end{figure*}

We introduce the principle of frequency modulation and demodulation of a timestream, a series of frequency-modulated spectra, as illustrated in figure~\ref{fig:principle-1} and \ref{fig:principle-2}.
We develop mathematical operations for frequency modulation and demodulation for both the signal and image sidebands.
We then introduce a reduction for generating a cleaned timestream that corresponds to $T\msb{A}^{\ast}$ (antenna temperature corrected for atmospheric absorption and spillover loss) of the conventional PSW method:
We reveal how the signal and noise are characterized in an intensity-calibrated timestream and how correlated components are defined and removed from it to make a cleaned timestream.
Finally, we describe how we convert a cleaned timestream into a final spectrum (single-pointed observation) or a map cube (OTF mapping observation).

\subsection{Mathematical expression of timestreams}
\label{subs:mathematical-expression-of-timestreams}

We define a timestream as a matrix that represents frequency and time.
Although frequency and time are originally continuous, we generally obtain an output of a digital spectrometer as a timestream that has discrete $D$ channels with a frequency width of $\Delta\nu$ (total bandwidth of $D\,\Delta\nu$), and discrete $N$ spectra with a data dump duration of $\Delta t$ (total observation time of $N\,\Delta t$).
Thus, we can express an arbitrary timestream, $\bX$, as a matrix that has $D$ rows and $N$ columns:
\begin{equation}
  \bX \equiv \brw{X_{dn}},
\end{equation}
where $X_{dn}$ is a scalar element at the $d$-th row and $n$-th column corresponding to the $d$-th spectrometer's channel at the $n$-th sampled spectrum.
In an observation with a heterodyne receiver, each row of $\bX$ should correspond to an intermediate frequency (IF; $\nu\msp{IF}$).
Without the FMLO, it exactly corresponds to an observed (radio) frequency (RF; $\nu\msp{RF}$) by a fixed LO frequency, $\nu\msp{LO}$:
\begin{equation}
  \nu\msp{RF} =
  \left\{
    \begin{array}{ll}
      \nu\msp{LO} + \nu\msp{IF} & (\mr{upper\ sideband})\\
      \nu\msp{LO} - \nu\msp{IF} & (\mr{lower\ sideband}).
    \end{array}
  \right.
  \label{eqn:rf-lo-if-relation-nonfm}
\end{equation}

We also define the mathematical operations between timestreams.
As there are many element-wise operations between two timestreams, we express them as scalar ones:
\begin{equation}
  \bX\bY \equiv \brw{X_{dn}Y_{dn}},\,\,
  \frac{\bX}{\bY} \equiv \brw{\frac{X_{dn}}{Y_{dn}}},\,\,
  \bX^{\bY} \equiv \brw{X_{dn}^{Y_{dn}}},
\end{equation}
where $X_{dn}$ and $Y_{dn}$ are elements of $\bX$ and $\bY$, respectively.
On the other hand, we explicitly express a normal matrix product using an at-sign operator.
For example, a matrix product of a $N \times D$ matrix, $\bX$, and a $D \times M$ matrix, $\bY$, is expressed as:
\begin{equation}
  \bX \,@\, \bY \equiv \brw{\sum_{d=1}^{D} X_{nd}Y_{dm}}.
\end{equation}
For convenience, we use bold symbols such as $\bzero$, $\bone$, and $\be$, which are $D \times N$ matrices filled with $0$, $1$, and $e$, respectively.

\subsection{Modulation and demodulation of timestreams}
\label{subs:modulation-and-demodulation-of-timestreams}

\begin{figure*}[t]
  \centering
  \includegraphics[width=\linewidth]{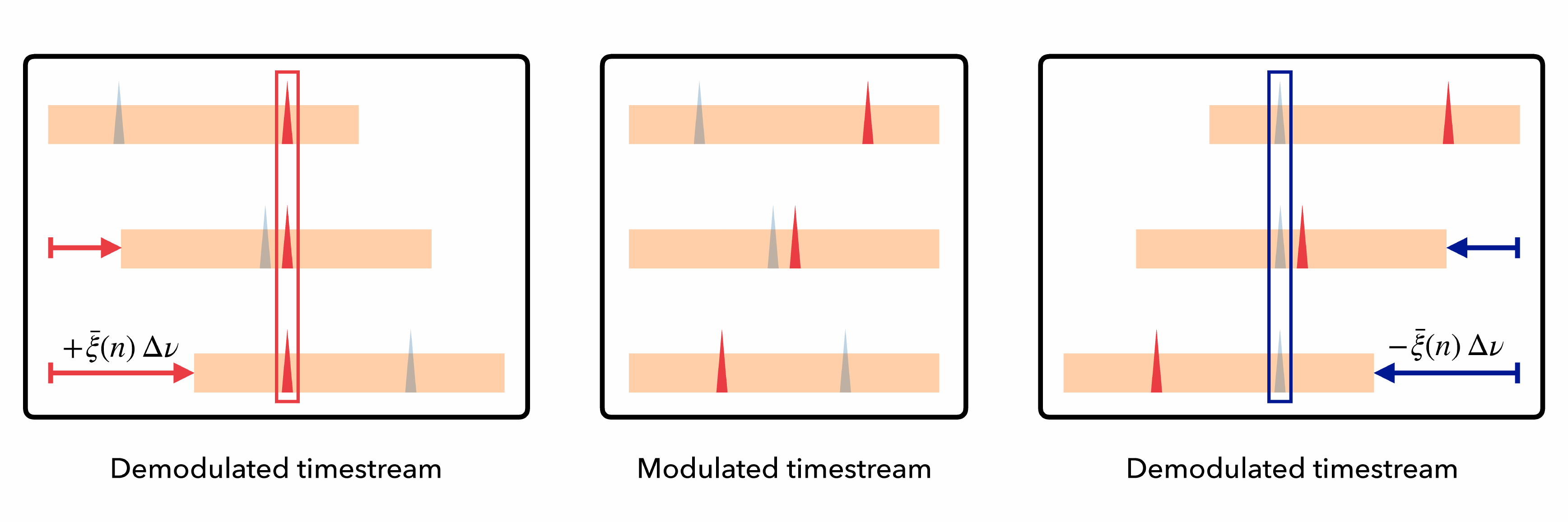}
  \caption{
    Schematic diagram of reverse-demodulation of a timestream.
    As we see in figure~\ref{fig:principle-1}, demodulation according to FM channels with their signs reversed will align a timestream with the RF frequency of the \emph{image} sideband:
    Such contamination can be modeled and subtracted if we integrate the reverse-demodulated timestream to generate a spectrum of contamination.
  }
  \label{fig:principle-2}
\end{figure*}

We define the frequency modulation as discrete changes in the LO frequency synchronized with data integration of a spectrometer.
As illustrated in figure~\ref{fig:principle-1}, we express the LO frequency, $\nu\msp{LO}(n)$, as the sum of a fixed LO frequency, $\nu\msp{LO,0}$, and a frequency offset from it as a function of time, $\Delta\nu\msp{LO}(n)$:
\begin{equation}
\nu\msp{LO}(n) = \nu\msp{LO,0} + \Delta\nu\msp{LO}(n).
\end{equation}
The observed frequency corresponding to an IF frequency (and also to a row of a timestream) is now time-dependent too:
\begin{equation}
  \nu\msp{RF}(n) =
  \left\{
    \begin{array}{ll}
      \nu\msp{LO}(n) + \nu\msp{IF} & (\mr{upper\ sideband})\\
      \nu\msp{LO}(n) - \nu\msp{IF} & (\mr{lower\ sideband}).
    \end{array}
  \right.
  \label{eqn:rf-lo-if-relation}
\end{equation}

We hereafter refer to $\Delta\nu\msp{LO}(n)$ as a frequency modulation pattern (an FM pattern).
It is an $N$-length vector and a new observational parameter to be determined by a user for an FMLO observation.
For the following mathematical computations, each value should be a multiple of $\Delta\nu$:
\begin{equation}
  \Delta\nu\msp{LO}(n) = \xi(n)\,\Delta\nu,
\end{equation}
where $\xi(n)$ is an integer that represents a channel-based FM pattern.
We hereafter refer to $\xi(n)$ as a frequency modulation channel (an FM channel).
For convenience, we also define the zero-based indexing FM channel, $\bar{\xi}(n)$, whose minimum value is zero by definition:
\begin{equation}
  \bar{\xi}(n) \equiv \xi(n) - \mr{min}(\brw{\xi(1),\dots,\xi(N)}).
  \label{eqn:blfmch}
\end{equation}

Now we express a modulated timestream with a $D \times N$ matrix, $\bX$, where each row of the timestream corresponds to an IF frequency.
Similarly, we express a demodulated timestream with a $\dm{D} \times N$ matrix, $\dm{\bX}$, where $\dm{D}$ is the number of spectrometer channels that cover the total observed width of the RF frequency\footnote{Hereafter, a symbol with tilde denotes a demodulated variable.}:
\begin{equation}
  \dm{D} \equiv D + \mr{max}(\brw{\bar{\xi}(1),\dots,\bar{\xi}(N)}).
\end{equation}
Finally, demodulation of $\bX$ and modulation of $\dm{\bX}$ can be expressed as $\dm{\bX} = \mc{M}^{-1}(\bX)$ and $\bX = \mc{M}(\dm{\bX})$, respectively, where $\mc{M}^{-1}: \bX \rightarrow \dm{\bX}$ and $\mc{M}: \dm{\bX} \rightarrow \bX$ are mapping operators defined by the following equations:
\begin{eqnarray}
  \dm{X}_{dn} &\leftarrow&
  \left\{
    \begin{array}{ll}
      X_{d-\bar{\xi}(n),\,n} & (d-\bar{\xi}(n)>0)\\
      \mr{NaN} & (\mr{otherwise}),
    \end{array}
  \right.\\
  X_{dn} &\leftarrow& \dm{X}_{d+\bar{\xi}(n),\,n}.
\end{eqnarray}
By definition, NaN (not a number) in $\dm{\bX}$ is not mapped to $\bX$, which guarantees that $\bX = \mc{M}\brp{\mc{M}^{-1}(\bX)}$ and $\dm{\bX} = \mc{M}^{-1}\brp{\mc{M}(\dm{\bX})}$.

\subsection{Sideband separation with reverse-demodulation}
\label{subs:sideband-separation-with-reverse-demodulation}

One of the advantages of the FMLO method is the software-based sideband separation in an offline data reduction achieved by modeling and subtracting the leaked line emission from an image sideband independently of that in the signal sideband.
This can reduce the noise induced by the leaked signal and improve the image sideband rejection ratio in an FMLO observation.
As illustrated in figure~\ref{fig:principle-2}, the IF frequency (i.e., spectrometer channel) corresponding to a fixed RF frequency in the upper sideband, $\nu\msp{RF}$, can be expressed as a function of the FM pattern:
\begin{equation}
  \nu\msp{IF}(n) = \nu\msp{RF} - \nu\msp{LO}(n) = -\Delta\nu\msp{LO}(n) + (\nu\msp{RF} - \nu\msp{LO,0}).
\end{equation}
Similarly, the IF frequency corresponding to a fixed RF frequency in the image sideband can be expressed as follows, however, the sign of the FM pattern is inverted:
\begin{equation}
  \nu\msp{IF}(n) = \nu\msp{LO}(n) - \nu\msp{RF,i} = +\Delta\nu\msp{LO}(n) + (\nu\msp{LO,0} - \nu\msp{RF,i}).
\end{equation}
This indicates that leaked signals from the image sideband are modulated reversely:
They can be modeled and subtracted when a timestream is reverse-demodulated by adopting $-\xi(n)$ as the FM channel instead of $+\xi(n)$, while the native signal is not (smeared out in a final product).

\subsection{Observation equation}
\label{subs:observation-equation}

Here we describe how signal and noise components are characterized in a timestream for making a cleaned timestream.
We can express a timestream of an on-point measurement after absolute intensity calibration, $\bT\msp{cal}$, as the sum of contributions from antenna temperatures in two sidebands, noise from the sky, and noise from instruments:
\begin{eqnarray}
\bT\msp{cal}
  &=& \bT\msp{a\ast}\exp(-\btau)
      + T\msb{atm}(\bone-\exp(-\btau))
      \nonumber\\
  &+& R\brq{\bT\msp{a\ast,i}\exp(-\btau\msp{i})
            + T\msb{atm}(\bone-\exp(-\btau\msp{i}))}
            \nonumber\\
  &+& \bE,
  \label{eqn:calibrated-timestream}
\end{eqnarray}
where $\bT\msp{a\ast}$ is a modulated antenna temperature of astronomical signals corrected for atmospheric absorption and spillover loss, $T\msb{atm}$ is the physical temperature of the sky, $\btau$ is the modulated opacity of atmosphere, $R$ is an image rejection ratio of a sideband separation mixer ($R=1$ for a double sideband mixer), and $\bE$ is noise attributed to the sky and instruments.
Symbols with i as the superscript express contributions from the image sideband\footnote{Hereafter, a symbol with a superscript of i denotes a variable of the image sideband.}.
We can decompose the following components into correlated and non-correlated ones\footnote{Hereafter, an equations like $\bX\msp{(,i)}=\bY\msp{(,i)}+\bZ\msp{(,i)}$ bundles two equations of the signal and image sidebands (i.e., $\bX=\bY+\bZ$ and $\bX\msp{i}=\bY\msp{i}+\bZ\msp{i}$, respectively).}:
\begin{eqnarray}
  \bT\msp{a\ast(,i)} &=& \bT\msp{c(,i)} + \bT\msp{nc(,i)},\\
  \btau\msp{(i)} &=& \btau\msp{c(,i)} + \btau\msp{nc(,i)},\\
  \bE &=& \bE\msp{c} + \bE\msp{nc}.
  \label{eqn:noise-components}
\end{eqnarray}
Hereafter, ``c'' and ``nc'' denote correlated and non-correlated timestreams, respectively.
The correlated components of $\bT\msp{c}$ and $\btau\msp{c}$ are attributed to continuum emission from astronomical signals and the sky, the latter of which usually fluctuates during an observation.
The non-correlated components of $\bT\msp{nc}$ and $\btau\msp{nc}$ are attributed to spectral line emission and/or absorption from astronomical signals and the sky (e.g., atmospheric ozone), respectively.
$\bE\msp{c}$ represents correlated noise, which is mainly attributed to the fluctuation in the bandpass gain coupled with the sky and instruments.
$\bE\msp{nc}$ represents the residual non-correlated noise that is expected to follow a multivariate Gaussian distribution, $\mc{N}(\bzero, \bI_{D})$, where $\bI_{D}$ is a $D \times D$ identity matrix.
Applying the correlated component removal method to $\bT\msp{cal}$ (see section~\ref{subs:correlated-component-removal} for more details), the entire correlated component, $\bT\msp{cor}$, can be estimated as the sum of terms in equation~\ref{eqn:calibrated-timestream}, which have at least one correlated component ($\bT\msp{c}\exp(-\btau\msp{c})$, for example).
Now, we rewrite equation~\ref{eqn:calibrated-timestream} using $\bT\msp{cor}$ and components of the line emission, which can be estimated separately:
\begin{equation}
  \bT\msp{cal}
  = \bT\msp{cor} + \bT\msp{ast} + \bT\msp{atm}
  + R\brp{\bT\msp{ast,i} + \bT\msp{atm,i}}
  + \bE\msp{nc},
  \label{eqn:all-components}
\end{equation}
where $\bT\msp{ast(,i)}$ and $\bT\msp{atm(,i)}$ represent modulated timestreams of astronomical and atmospheric line emissions, respectively:
\begin{eqnarray}
  \bT\msp{ast(,i)} &\equiv& \bT\msp{nc(,i)}\exp(-\btau\msp{nc(,i)}),\\
  \bT\msp{atm(,i)} &\equiv& T\msb{atm}(\bone-\exp(-\btau\msp{nc(,i)})).
\end{eqnarray}
After estimating and subtracting components other than $\bT\msp{ast}$, we finally obtain a modulated cleaned timestream composed of astronomical signals of interest, $\bT\msp{cln}$, corresponding to the so-called $T\msb{A}^{\ast}$ of the conventional PSW method:
\begin{eqnarray}
  \bT\msp{cln}
  &\equiv& \bT\msp{cal} - \bT\msp{cor} - \bT\msp{atm} - R\brp{\bT\msp{ast,i} + \bT\msp{atm,i}}\nonumber\\
  &\simeq& \bT\msp{nc}\exp(-\btau\msp{nc}) + \bE\msp{nc}.
  \label{eqn:cleaned-timestream}
\end{eqnarray}
If the spectral line emission from astronomical signals does not overlap with those from the sky, the equation~\ref{eqn:cleaned-timestream} is simply expressed as
\begin{equation}
  \bT\msp{cln}
  \simeq \bT\msp{nc} + \bE\msp{nc}.
\end{equation}
Otherwise, the contribution of the line emission from the sky, $\btau\msp{nc}$, should be derived from $\bT\msp{atm}$ and used for the correction of equation~\ref{eqn:cleaned-timestream}.

\subsection{Correlated component removal}
\label{subs:correlated-component-removal}

We estimate the entire correlated component, $\bT\msp{cor}$, by principal component analysis (PCA).
It is originally an orthogonal matrix transformation that converts a $D \times N$ correlated matrix, $\bX$ (mean values are assumed to be subtracted), into a linearly non-correlated one, $\bC$:
\begin{eqnarray}
  \bX = \bP \,@\, \bC
  \quad\Leftrightarrow\quad
  \bC = \bP^{T} \,@\, \bX,
\end{eqnarray}
where $\bP$ is a $D \times \mr{min}(D,N)$ transformation matrix composed of the eigenvectors of the covariance matrix (i.e., $N^{-1}\bX^{T}\bX$).
$\bC$ is a $\mr{min}(D,N) \times N$ matrix named the principal component matrix, because it is defined such that the first principal component has the largest variance and subsequent ones have the second, third, $\dots$, largest variances, and are orthogonal to the other components.
PCA is widely used to, for example, extract features of data with fewer ($< D$) variables or visualize high-dimensional data as a two- or three-dimensional plots \citep{Jolliffe2002}.
From the viewpoint of correlated component removal, PCA is an effective method for estimating such components, because it is a low-rank approximation methods of a matrix.
Correlated components, $\bX\msp{c}$, can be modeled as a reconstruction of $\bX$ with only $K (<\mr{min}(D,N))$ largest principal components and eigenvectors:
\begin{equation}
  \bX\msp{c} \simeq \bP_{:,:K} \,@\, \bC_{:K,:},
  \label{eqn:correlated-components}
\end{equation}
where $\bP_{:,:K}$ is a $D \times K$ matrix of the $K$ largest eigenvectors and $\bC_{:K,:}$ is a $K \times N$ matrix of the corresponding principal components.
As non-correlated components, $\bX\msp{nc}$, are expected to have smaller and uniform variances in the $D$-dimensional space, they shall remain with the rest of the principal components:
\begin{equation}
  \bX\msp{nc} \simeq \bX - \bX\msp{c}.
\end{equation}

\subsection{Making final product}
\label{subs:making-final-product}

\begin{figure*}[t]
  \centering
  \includegraphics[width=\linewidth]{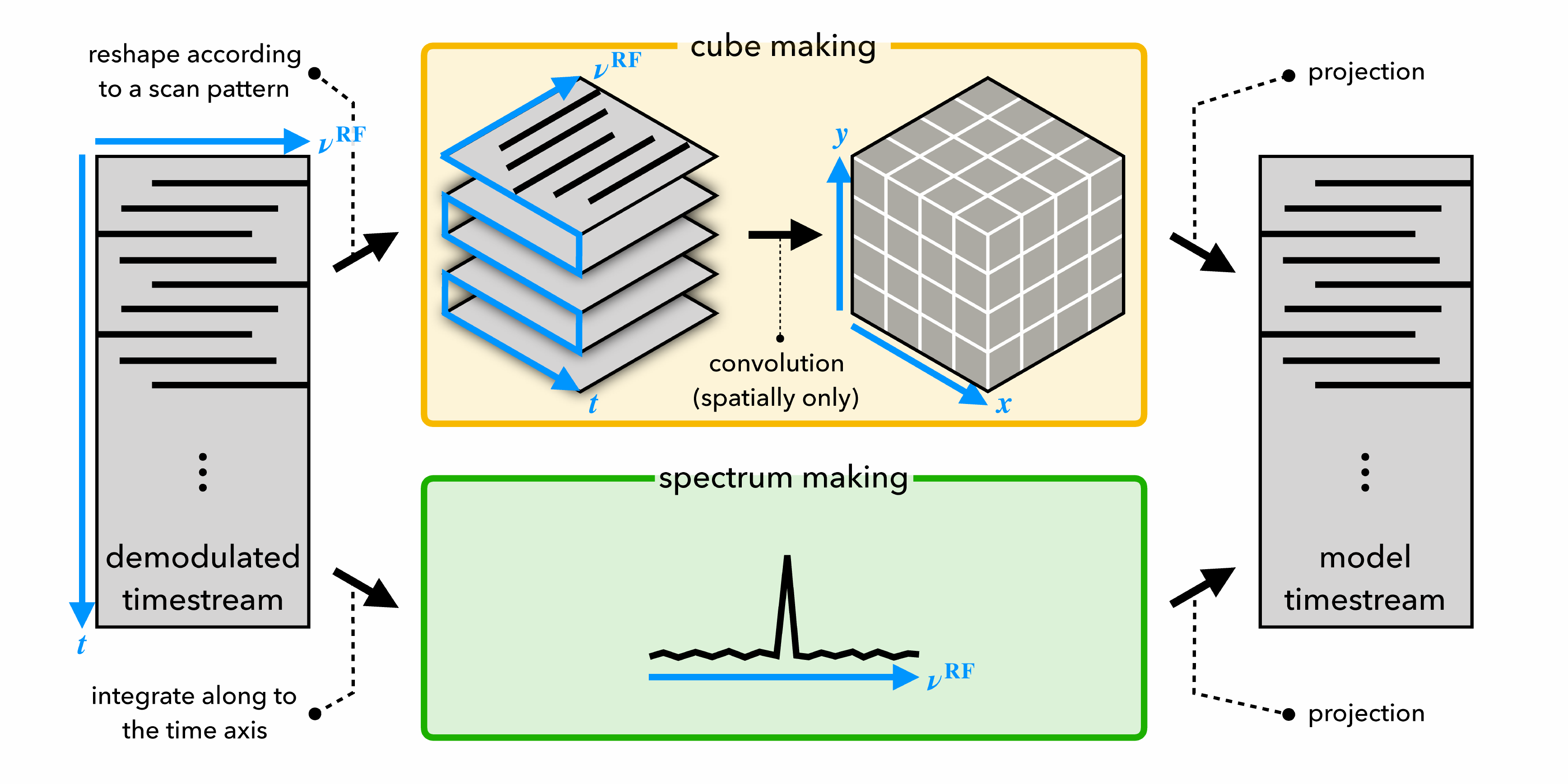}
  \caption{
    Schematic diagram of spectrum and map making.
    (left) A demodulated timestream whose non-NaN values are expressed as black line segments.
    (top center) A reshaped (folded) 3D-cube-like timestream according to a scan pattern of a mapping observation.
    Then the spatial convolution process converts the timestream into a 3D map cube.
    (bottom center) A spectrum derived from the timestream integrated along the time axis.
    (right) A demodulated model timestream of signal.
    In the case of a spectrum, a projection process generates a demodulated timestream in which each time sample is filled with a spectrum; the demodulated timestream is used to generate a modulated timestream.
    In the case of a map, a projection process converts a 3D map cube into a timestream whose shape is the same as that of the input timestream, which can be derived by 2D (spatial axes) interpolation of a 3D map cube at map coordinates at each observed time.
  }
  \label{fig:principle-3}
\end{figure*}

Once the cleaned timestream, $\bT\msp{cln}$, is obtained, a spectrum or a map is obtained for single-pointed or mapping observation, respectively, by demodulating $\bT\msp{cln}$, i.e., $\dm{\bT}\msp{cln} = \mc{M}^{-1}(\bT\msp{cln})$.
As illustrated in figure~\ref{fig:principle-3}, the methods of making such final products are the same as those of PSW or OTF mapping observations except that they contain NaNs in an obtained $\bT\msp{cln}$; they must be excluded when a spectrum or map is made.
This means that the total on-source time per RF channel is not constant over the observed band but is a function of the FM pattern, $\nu\msp{LO}(n)$.
Now, we define a $\dm{D} \times N$ weight matrix, $\dm{\bW}\msp{NaN}$, in order to handle such NaNs and thus the dependency of the FM pattern:
\begin{eqnarray}
  \dm{W}\msp{NaN}_{dn} =
  \left\{
    \begin{array}{ll}
      1 & (\dm{T}_{dn} \neq \mr{NaN})\\
      0 & (\dm{T}_{dn} = \mr{NaN}).
    \end{array}
  \right.
\end{eqnarray}

A spectrum of an FMLO observation is defined as a $\dm{D}$-length vector, $\dm{\bs}$, which is simply derived from the mean of $\bT\msp{cln}$ along the time axis excluding NaNs:
\begin{equation}
  \dm{s}_{d}
  = \bra{\dm{\bT}; \dm{\bW}\msp{NaN}}_{n}
  \equiv \frac
    {\sum_{n=1}^{N} \dm{W}\msp{NaN}_{dn} \, \dm{T}\msp{cln}_{dn}}
    {\sum_{n=1}^{N} \dm{W}\msp{NaN}_{dn}},
\end{equation}
where $\bra{X_{dn}; W_{dn}}_{n}$ represents the weighted mean of the $d$-th row of $\bX$ along the time axis with a weight of $\bW$.
The total on-source time is also defined as a $\dm{D}$-length vector, $\dm{\bt}$:
\begin{equation}
  \dm{t}_{d}
  = \sum_{n=1}^{N} \dm{W}\msp{NaN}_{dn} \cdot \brp{\eta\msb{obs} \, \Delta t}.
\end{equation}

A map of an FMLO observation can be defined as a $N_{x} \times N_{y} \times \dm{D}$ tensor, $\dm{\bM}$ (i.e., three-dimensional data cube).
$N_{x}$ and $N_{y}$ are the horizontal and vertical numbers of grids of a map, respectively, which depend on the coordinate system, the mapping area, and the grid spacing coupled with a target and the HPBW of a telescope.
Spectra at each map grid are derived from the weighted mean of samples that are obtained within a certain radius from the grid coordinate.
According to \citet{Sawada2008}, weight values are calculated by a gridding convolution function (GCF), $c(r)$, where $r$ is the distance between the antenna coordinates of a sample and a grid coordinates in units of grid spacing.
For example, the pure Gaussian GCF can be expressed as the following equation:
\begin{eqnarray}
  c(r) =
  \left\{
    \begin{array}{ll}
      \exp(-r^{2}) & (r \leq r\msb{max})\\
      0 & (\mr{otherwise}),
    \end{array}
  \right.
\end{eqnarray}
where $r\msb{max}$ represents the maximum radius within which samples are counted to calculate the weighted mean.
If we express $r$ regarding the $n$-th sample and grid $(x,y)$ as $r_{xyn}$, a $N_{x} \times N_{y} \times N$ weight tensor, $\dm{\bW}\msp{GCF}$, is expressed as
\begin{equation}
  \dm{W}\msp{GCF}_{xyn}
  = c(r_{xyn}).
\end{equation}
Finally a map, $\dm{\bM}$, and the total on-source time per RF channel at a grid, $\dm{\bt}$, are expressed as
\begin{eqnarray}
  \dm{M}_{xyd}
  &=& \bra{\dm{\bT}; \dm{\bW}\msp{GCF}, \dm{\bW}\msp{NaN}}_{n}
  \equiv \frac
    {\sum_{n} \dm{W}\msp{GCF}_{xyn} \, \dm{W}\msp{NaN}_{dn} \, \dm{T}\msp{cln}_{dn}}
    {\sum_{n} \dm{W}\msp{GCF}_{xyn} \, \dm{W}\msp{NaN}_{dn}},\\
  \dm{t}_{xyd}
  &=& \sum_{n=1}^{N} \dm{W}\msp{GCF}_{xyn} \, \dm{W}\msp{NaN}_{dn} \cdot \brp{\eta\msb{obs}\,\Delta t}.
\end{eqnarray}

\subsection{Making a model timestream from a final product}
\label{subs:Making-a-model-timestream-from-a-final-product}

A final product needs to be modeled to make a noise-free spectrum or map, and then the model product is transformed to make a noise-free timestream of astronomical signals, which is used in the iterative pipeline algorithm described in section~\ref{subs:data-reduction-procedure}.
We use the $\sigma$-cutoff method to make a noise-free product:
\begin{eqnarray}
  \dm{s}\msp{\,\,model}_{d} &=&
  \left\{
    \begin{array}{ll}
      \dm{s}_{d} & (|\dm{s}_{d}| > \theta\msb{cutoff} \, \dm{\sigma}_{d})\\
      0 & (\mr{otherwise}),
    \end{array}
  \right.\\
  \dm{M}\msp{model}_{xyd} &=&
  \left\{
    \begin{array}{ll}
      \dm{M}_{xyd} & (|\dm{M}_{xyd}| > \theta\msb{cutoff} \, \dm{\sigma}_{xyd})\\
      0 & (\mr{otherwise}),
    \end{array}
  \right.
\end{eqnarray}
where $\dm{\bsigma}$ is the standard deviation of $\dm{\bs}$ or $\dm{\bM}$, which is derived by weighted means, i.e., $\sqrt{\bra{X^{2};W}_{n}-\bra{X;W}_{n}^{2}}$, and $\theta\msb{cutoff}$ is a threshold signal-to-noise ratio.

As illustrated in figure~\ref{fig:principle-3}, the model product is transformed into a demodulated model timestream, $\dm{\bT}\msp{model}$.
In the case of a spectrum, it is a $\dm{D} \times N$ matrix whose columns are filled with $\dm{\bs}\msp{\,\,model}$.
In the case of a map, it is a $\dm{D} \times N$ matrix whose $n$-th column is a spectrum as a result of two-dimensional interpolation (i.e., x and y axes) of $\dm{\bM}\msp{model}$ at its antenna coordinates.

\section{Instrumentation}
\label{s:instrumentation}

\subsection{Hardware implementation}
\label{subs:hardware-implementation}

\begin{figure*}[t]
  \centering
  \includegraphics[width=\linewidth]{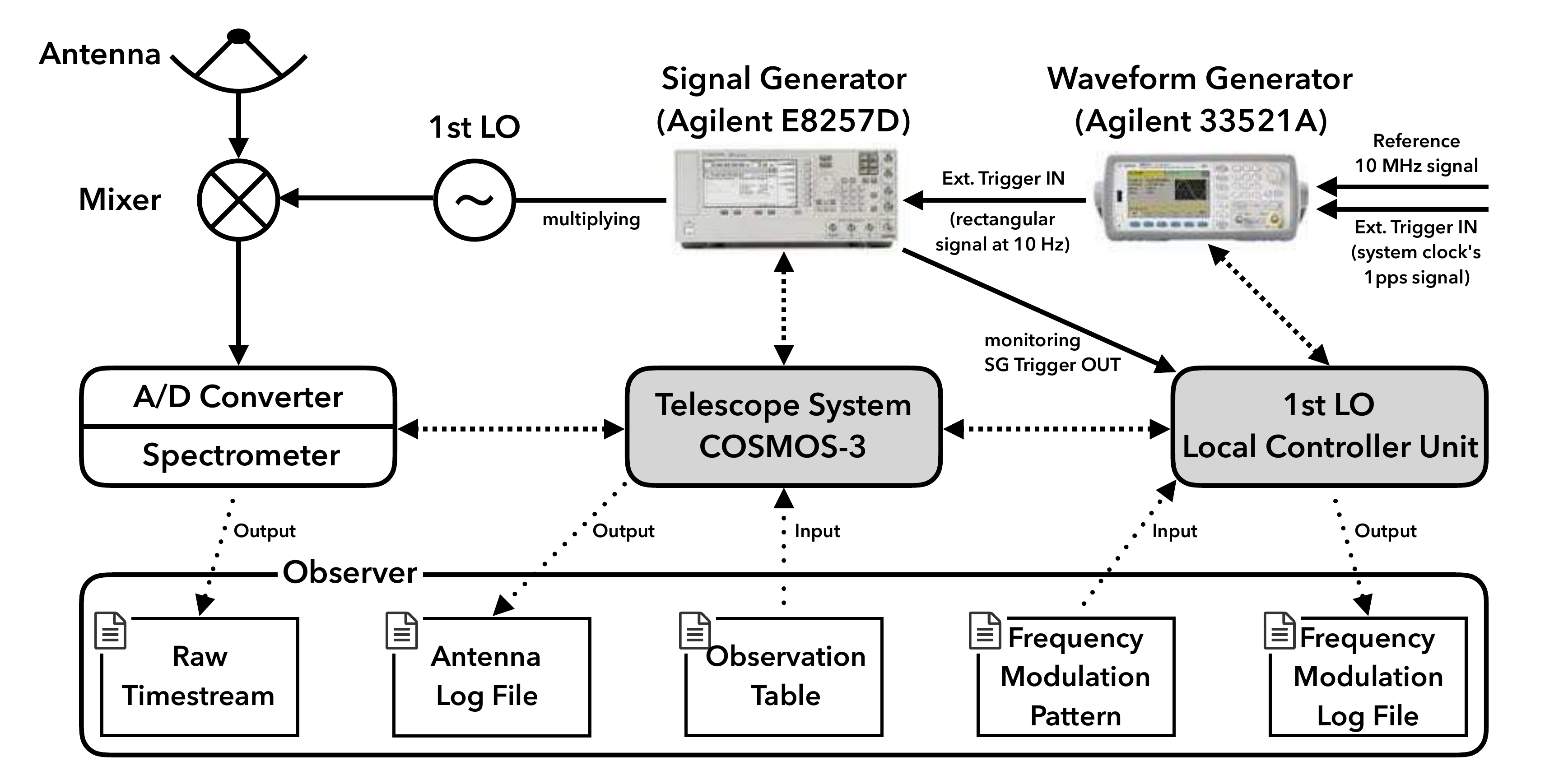}
  \caption{
    Block diagram of an FMLO system on the Nobeyama 45-m.
    The solid and dashed arrows indicate the directions of signals and data communications between instruments and an observer, respectively.
    The diagram has three layers: (top) the frontend receiver system, where the RF signal from the sky and frequency-modulated LO signal are mixed at the SIS device and a subsequent IF signal is input to the spectrometer after analog-to-digital conversion; (middle) the backend spectrometer and the telescope system, \texttt{COSMOS-3}; (bottom) an observer who sends and receives inputs and outputs.
  }
  \label{fig:blockdiagram}
\end{figure*}

In the proposed FMLO method, it is essential to modulate astronomical signals of interest into high time frequency ranges in a timestream by modulating the observing frequency, which allows for the isolation of astronomical signals from correlated noise of a low time frequency.
Although there are several methods to modulate the frequency, we choose to modulate radio frequency (RF) signals.
This is because (1) in many modern systems, a first LO is realized with a computer-controlled signal generator in which a built-in modulation function is implemented and (2) RF modulation in mm/submm allows for a wide (GHz-order) frequency change compared with IF modulation.

The minimum requisites for the telescope system on which the FMLO system is to be installed are as follows: (1) a tunable and programmable first LO; (2) a system clock that ensures synchronization between frequency modulation and data acquisition; and (3) a backend spectrometer that takes the data at a dump rate sufficiently higher than variations in the sky and system.
A heterodyne receiver in modern mm/submm astronomy often utilizes a microwave signal generator with a cascade of frequency multipliers, instead of a Gunn oscillator, as a first LO.
A digital signal generator is particularly useful for the purpose of the FMLO method, as it is easy to quickly tune the LO frequency and program the FM pattern.
A dump rate of $\sim$10~Hz should be sufficient for many cases; the time-scale of sky variation is of the order of $\sim$1~s, as it is roughly determined by the crossing-time in which the phase screen ($v \sim 10$ m\,s$^{-1}$) goes across the telescope aperture ($D \sim 10$~m).
Note that when we apply the FMLO method to on-the-fly (OTF) mapping rather than single-pointed observations, synchronization between frequency modulation and antenna drive control is also required.

\begin{figure*}[t]
  \centering
  \includegraphics[width=\linewidth]{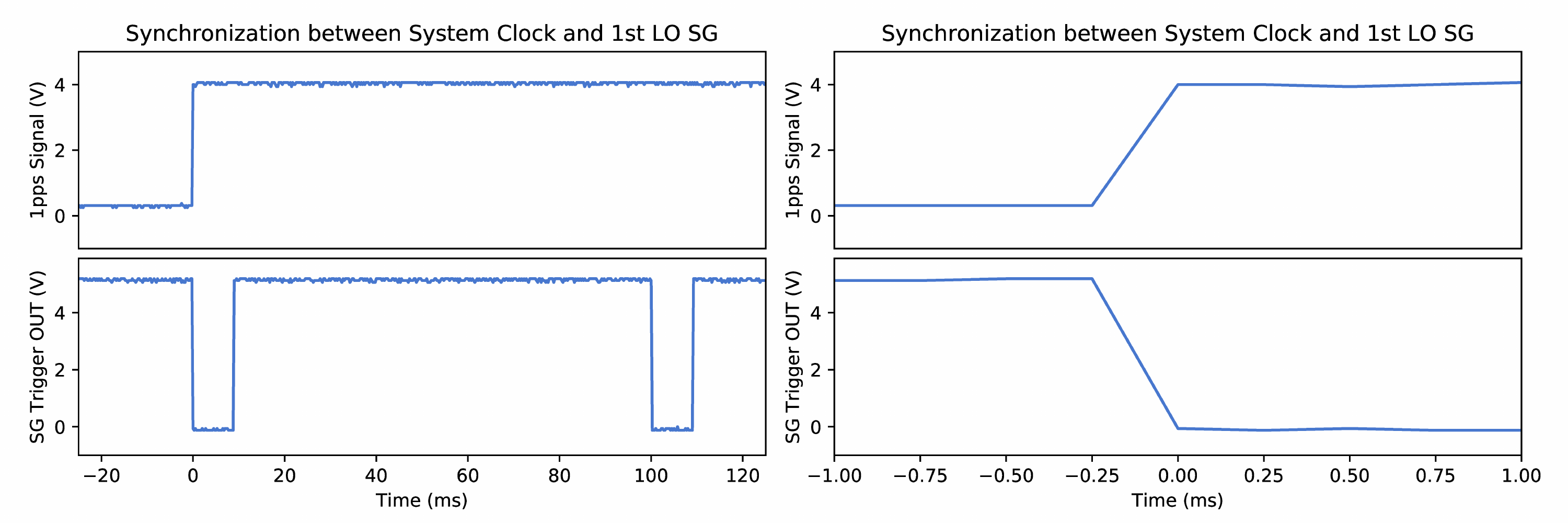}
  \caption{
    The measured signals of the 1-pps system clock (top) and 1st LO signal generator's reference trigger (bottom) of the Nobeyama 45-m telescope in units of voltage.
    The left panels show them over a $\Delta t = 150$~ms duration, where the 1-pps signal rises at $t=0$~ms, while the trigger signal, which is synchronized with the 1-pps clock, falls.
    The subsequent $\Delta t \simeq 8$~ms voltage dropping at $0$~V is attributed to the settling time of the signal generator, where it does not generate a signal for the LO; thus, the SIS mixer is unavailable.
    The right panels show the same results, but over a $\Delta t = 2$~ms duration around $t=0$~ms, which demonstrates that the time synchronization error is much better than $200~\mu$s, the period of a slope.
  }
  \label{fig:timesync}
\end{figure*}

As an example of hardware implementation, we show a block diagram of an FMLO observing system on the Nobeyama 45-m (developed in 2013) in figure~\ref{fig:blockdiagram}.
The receiver system comprises the two-beam TZ front-end receiver with a cryogenic superconductor-insulator-superconductor (SIS) mixer \citep{Nakajima2013} and the digital backend spectrometer SAM45, which is an exact copy of the ALMA ACA correlator \citep{Kamazaki2012}. In this study, we just uses a single IF (a single beam and a single polarization).
We employ a signal generator, Agilent E8257D, which is capable of generating a continuous wave (CW) according to a frequency list given by an observer.
Each frequency in the list is switched to by external TTL-compatible reference triggers in a sequential manner.
The trigger is produced with an arbitrary waveform generator, Agilent E33521A.
The waveform generator produces a rectangular wave with a period of 100~ms, which is synchronized with the telescope' s system clock via 1~pps and 10~MHz reference signals.
The period must be identical to the dump rate of the spectrometer outputs (10~Hz), and the phase of the rectangular wave must be synchronized with the onset of data acquisition.
This is made in the on-the-fly (OTF) mode of the Nobeyama 45-m \citep{Sawada2008}, while the telescope may focus on a single point in the sky.
Figure~\ref{fig:timesync} shows the voltages of the reference trigger and 1-pps signal as a function of time, which shows accurate enough synchronization.
The typical error in synchronization is better than $\simeq 200~\mu$s, which is well below the typical dwell time of a single frequency (100~ms).
Note that it typically takes $\lesssim 8$~ms to settle the generated LO frequency after the frequency is set to one value from another.
The Agilent E8257D does not output any CW signals during the interval in which the frequency settles to a programmed value, which makes the SIS device deactivate itself temporally, and thus the SIS is unavailable during the settling time.
This causes a slight sensitivity loss of $\lesssim 4\%$ for a dwell time of 100~ms (i.e., $((100-8)/100)^{1/2} \simeq 0.96$).
The decrease in astronomical signals is corrected for in an absolute intensity calibration.

A typical procedure of an FMLO observation of the Nobeyama 45-m is as follows.
The data and signal flow are shown at the bottom of figure~\ref{fig:blockdiagram}.
\begin{enumerate}
\item
The Nobeyama 45-m telescope system (\texttt{COSMOS-3}; \cite{Morita2003, Kamazaki2005}) loads a script for an FMLO observation (observation table); then the local controller unit (LCU) reads a frequency list (FM pattern file) and sends it to the signal generator.
\item
Once the telescope begins stable tracking, the receiver is properly tuned, and at this time, the spectrometer is ready to record; then, the LCU triggers the signal generator when data acquisition starts.
\item
During an observation, the spectrometer records a time-series spectra of an on-point at a dump rate of 10~Hz while it regularly takes measurements of a hot load (chopper wheel) at the reference frequency for an absolute intensity calibration (typically once every 30~min).
At the same time, the LCU logs incident information about the frequency, timestamp, and Doppler tracking of the receiver into a frequency modulation log file.
\item
After the observation, an observer obtains a raw timestream, an antenna log file that contains time-series antenna coordinates, and a frequency modulation log file that contains actual time-series FM values generated by an LO.
\end{enumerate}

\subsection{Data reduction procedure}
\label{subs:data-reduction-procedure}

\begin{figure*}[t]
  \centering
  \includegraphics[width=\linewidth]{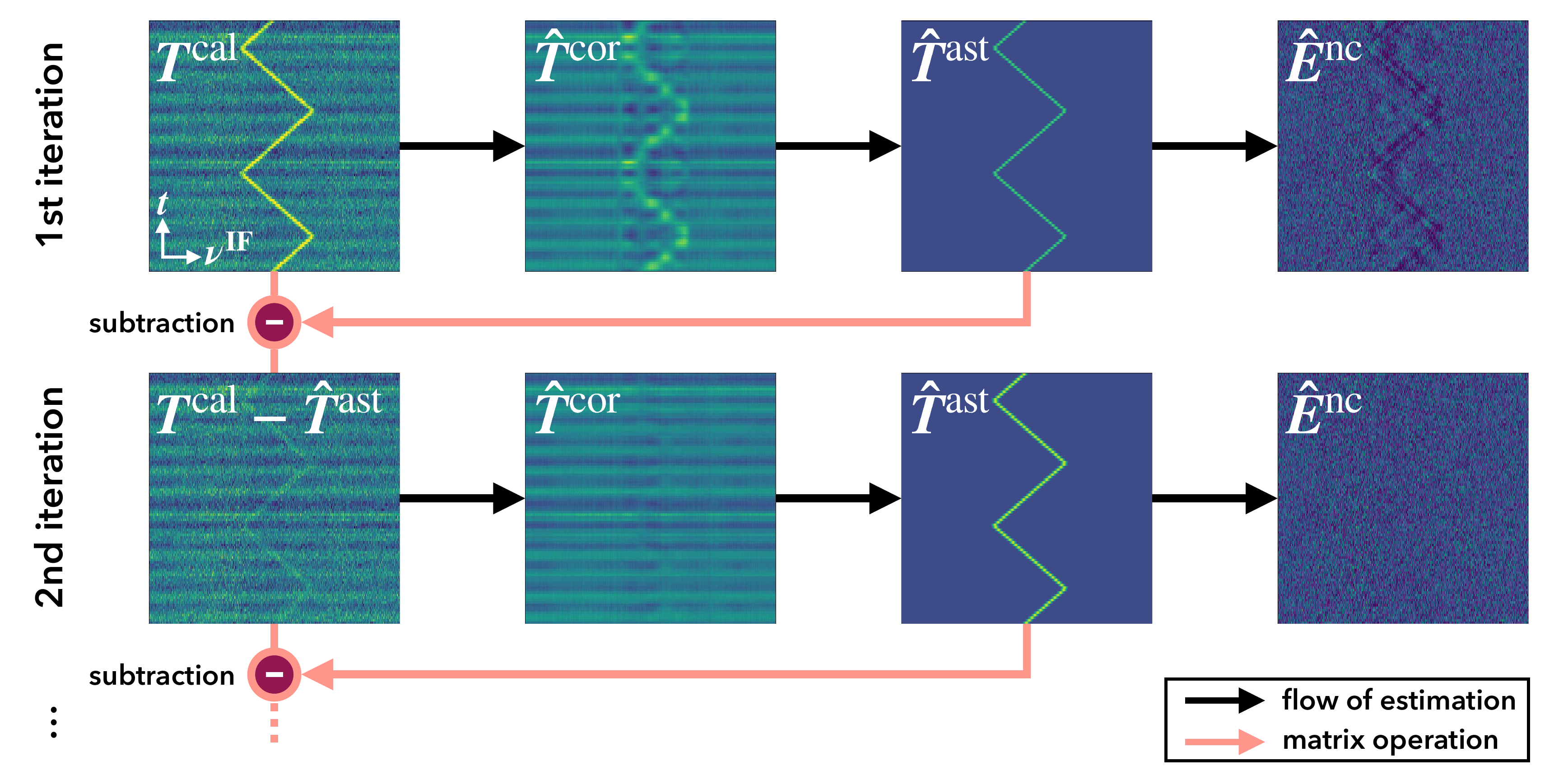}
  \caption{
    Flowchart of the iterative algorithm in an FMLO data reduction.
    For simplification, we show the case of no astronomical line emission from the image sideband (i.e., $\bT\msp{cal}=\bT\msp{cor}+\bT\msp{ast}+\bE\msp{nc}$ is assumed).
    Each panel represents a simulated modulated timestream observed with a zig-zag FM pattern (see also figure~\ref{fig:zigzag-fmp}).
    We assume that there exists a strong line emission observed around the center of the spectrometer band.
    The top left panel shows the timestream of the measured antenna temperature, $\bT\msp{cal}$.
    The other panels in the top row show the estimates of the correlated components ($\hat{\bT}\msp{cor}$), astronomical line emission ($\hat{\bT}\msp{ast}$), and residual ($\hat{\bE}\msp{nc}$) in the first iteration.
    The bottom panels show the estimates of the correlate components in the second iteration.
    The estimation starts from $\bT\msp{cal}$ with the subtraction of $\hat{\bT}\msp{ast}$, which results in better estimation of $\hat{\bT}\msp{cor}$.
  }
  \label{fig:flowchart}
\end{figure*}

After an FMLO observation, the succeeding data reduction is conducted offline to make a final product (a spectrum or map).
It is thus necessary to handle outputs properly and apply the signal processing methods to them according to equations described in section~\ref{s:principle}.
We merge such outputs into a single file, the format of which is independent of hardware implementation, and create and operate a two-dimensional array representing a modulated timestream of on-point spectra, which is loaded from the file in an offline pipeline program.
We choose to use \texttt{FITS} (flexible image transport system) as the file format and develop a Python-based data analysis package, \texttt{FMFlow}\footnote{\url{https://github.com/fmlo-dev/fmflow} (DOI 10.5281/zenodo.3433962)}.
It provides functions for timestream operations such as modulation and demodulation, correlated noise removal by PCA, and generating a final product from a timestream and vice versa (i.e., generating a model timestream from a final product).

In the data reduction process, it is also essential to implement an \emph{iterative} algorithm to estimate $\bT\msp{cor}$, $\bT\msp{ast(,i)}$, and $\bT\msp{atm(,i)}$ by turns.
The iterative method was originally introduced by an iterative map-making algorithm for the bolometer array camera, SCUBA-2 \citep{Chapin2013}, in which map-based correlated components (referred as common-mode) and astronomical signals were estimated.
On the other hand, our method estimates them based on a spectrum and optimizes them for sideband separation as mentioned in section~\ref{s:principle}.
We show a flowchart of the iterative algorithm in figure~\ref{fig:flowchart}.
With a single estimate of the spectrum-based correlated components, an estimate of correlated noise, $\hat{\bT}\msp{cor}$, might be strongly affected by the line emission from the sky and/or astronomical signals\footnote{Hereafter, a symbol with a hat denotes a variable of an estimate.}.
This usually yields negative sidelobe-like features around the line emission in the final spectrum.
Such features can also be seen in the residual timestream, $\hat{\bE}\msp{nc}$, at the top right panel in figure~\ref{fig:flowchart}.
It is therefore necessary to model\footnote{Here, ``model'' does not mean to obtain an estimate of the true value; rather, it means to make a best-effort and noise-free ones at each iteration used for the next one.} $\bT\msp{ast(,i)}$ and $\bT\msp{atm(,i)}$, and re-estimate $\bT\msp{cor}$ from a timestream where $\hat{\bT}\msp{ast(,i)}$ and $\hat{\bT}\msp{atm(,i)}$ are subtracted.
Such subsequent iterative processes can minimize any errors between an estimate and a ``true'' value as the estimate is converged after several iterations.

Here we introduce the actual algorithm.
For simplicity, we suppose an observed situation where the atmospheric line emission does not exist in the observed band:
\begin{equation}
  \bT\msp{cal}
  \simeq \bT\msp{cor} + \bT\msp{ast} + \bT\msp{ast,i} + \bE\msp{nc},
  \label{eqn:simple-components}
\end{equation}
where we express $R\,\bT\msp{ast,i}$ as $\bT\msp{ast,i}$ for simplicity, too.
The steps of the algorithm are as follows:
\begin{enumerate}
  \item Set initial estimates of $\bT\msp{ast(,i)}$ to zero ($\hat{\bT}\msp{ast}=\hat{\bT}\msp{ast,i}=\bzero$).
  \item Estimate correlated components, $\hat{\bT}\msp{cor}$, by applying PCA to the timestream of $\bT\msp{cal}-\hat{\bT}\msp{ast}-\hat{\bT}\msp{ast,i}$ (i.e., deriving $\bX\msp{c}$ in equation~\ref{eqn:correlated-components}, where $\bX=\bT\msp{cal}-\hat{\bT}\msp{ast}-\hat{\bT}\msp{ast,i}$).
  \item Estimate the astronomical line emission from the signal sideband, $\hat{\bT}\msp{ast}$, by modeling a timestream from the final product derived from $\bT\msp{cal}-\hat{\bT}\msp{cor}-\hat{\bT}\msp{ast,i}$.
  \item Estimate the astronomical line emission from the image sideband, $\hat{\bT}\msp{ast,i}$, by modeling a timestream from the final product derived from $\bT\msp{cal}-\hat{\bT}\msp{cor}-\hat{\bT}\msp{ast}$.
  \item Generate the cleaned timestream, $\hat{\bT}\msp{cln} = \bT\msp{cal}-\hat{\bT}\msp{cor}-\hat{\bT}\msp{ast,i}$. If $\hat{\bT}\msp{cln}$ is converged (i.e., values are not significantly changed from the previous ones), the iterative algorithm is finished. Otherwise, return to step 2 and repeat until convergence is achieved.
\end{enumerate}
The convergence of a matrix, $\bT\msp{new}$ compared to the previous one, $\bT\msp{old}$, is checked by determining whether the following condition is fulfilled or not:
\begin{equation}
  \left| \frac{\bT\msp{new}-\bT\msp{old}}{\bT\msp{old}} \right|_{F} < \eps,
\end{equation}
where $|\cdot|_{F}$ is a Frobenius norm\footnote{A Frobenius norm of a matrix, $|\bX|_{F}$, is defined as $|\bX|_{F} \equiv \brp{\sum_{d=1}^{D}\sum_{n=1}^{N}X_{dn}^{2}}^{1/2}$.}, and $\eps$ is a threshold value.
This means that a matrix, $\bT$, is regarded as having converged if the total variation in $\bT$ from the previous one is less than $\varepsilon$.

\section{Demonstration}
\label{s:demonstration}

We show the results of single-pointed and mapping observations with both the FMLO and PSW methods and demonstrate an improvement in observation efficiency of the FMLO and consistency of intensity between the two methods.
We use Galactic sources that have bright ($T\msb{A}^{\ast}\sim10^{0-1}$~K at peak) emission lines at millimeter wavelengths and thus are usually observed as ``standard sources'' for absolute intensity calibration in a spectral line observation.

\subsection{Observation efficiency and sensitivity improvement}
\label{subs:observation-efficiency-and-sensitivity-improvement}

We denote the $1\sigma$ noise level (sensitivity) of each spectrometer channel of a final spectrum as $\Delta T$.
In the conventional position switching methods, it is represented as the root sum of the noises from the on- and off-points:
\begin{equation}
  \Delta T^{2}
  = \Delta T\msb{on}^{2} + \Delta T\msb{off}^{2}
  = \frac{\alpha^{2}\, T\msb{sys}^{2}}{\Delta\nu\, t\msb{on}},
  \label{eqn:on-off-sensitivity}
\end{equation}
where $T\msb{sys}$ is the system noise temperature of a telescope at the observed frequency (expected to be constant during an observation), $\Delta \nu$ is the channel width of a spectrometer, $\alpha$ is a factor on the order of unity that represents the additional noise contribution from the off-point:
\begin{equation}
  \alpha\msb{PSW} = \sqrt{1 + \frac{t\msb{on}}{t\msb{off}}},
\end{equation}
where $t\msb{on}$ and $t\msb{off}$ are the on- and off-source integration times, respectively.
As they are often equal in a PSW observation (i.e., $\alpha\msb{PSW} = \sqrt{2}$), we can express $\Delta T\msb{PSW}$ as
\begin{equation}
  \Delta T\msb{PSW} = \frac{\sqrt{2}\, T\msb{sys}}{\sqrt{\Delta\nu\, t\msb{on}}}.
\end{equation}
On the other hand, the factor in an OTF mapping observation, $\alpha\msb{OTF}$, highly depends on observational parameters of a scan pattern.
We will derive it in section~\ref{subs:mapping-observation}.

In the proposed FMLO method, we do not observe the off-point but we do model it by the correlated component removal.
In this case, the additional noise contribution from the estimated off-point is only dependent on the accuracy of the correlated noise removal, which suggests that the noise is less than those of the PSW and OTF methods.
As $\Delta T$ of the FMLO method is also proportional to $T\msb{sys}/\sqrt{\Delta\nu\, t\msb{on}}$, we can write $\Delta T\msb{FMLO}$ as
\begin{equation}
  \Delta T\msb{FMLO} = \frac{\alpha\msb{FMLO}\, T\msb{sys}}{\sqrt{\Delta\nu\, t\msb{on}}},
  \label{eqn:fmlo-sensitivity}
\end{equation}
where $\alpha\msb{FMLO}$ is a factor of noise contribution from the model and is expected to be less than $\sqrt{2}$.
If comparing both noise levels obtained with the PSW and FMLO methods for a fixed on-source integration time, the FMLO method is expected to improve the sensitivity by a factor of $\sqrt{2}/\alpha$ compared with the PSW method.
If comparing both noise levels of the PSW and FMLO methods with the same \emph{total observation time}, which is a more practical situation in actual observations, the FMLO method is expected to improve sensitivity more because its observation efficiency ($\eta\msb{obs}$; equation~\ref{eqn:observation-efficiency}) is much higher than that of the PSW method.

In the ideal cases where the overhead time such as the telescope slue time between the on- and off-points is negligible, the observation efficiency of the PSW method is $\eta\msb{obs}\msp{PSW}\simeq 0.5$ because the on- and off-source integration times are equal.
On the other hand, the observation efficiency of FMLO is $\eta\msb{obs}\msp{FMLO}\simeq 0.92$ because of a settling time ($\lesssim 8$~ms) for each dump duration (100~ms) of the spectrometer (see section~\ref{subs:hardware-implementation}).
The sensitivity improvement of the FMLO method compared to that of the PSW method per unit total observation time, $\iota$, is thus expressed as the following equation:
\begin{equation}
    \iota
    = \frac{\sqrt{2}}{\alpha\msb{FMLO}}\brp{\frac{\eta\msb{obs}\msp{PSW}}{\eta\msb{obs}\msp{FMLO}}}^{-1/2}.
    \label{eqn:fmlo-sensitivity-improvement}
\end{equation}
This equation indicates that the FMLO observation requires only $1/\iota^{2}$ of total observation time compared to that of the PSW to achieve the same sensitivity of the final spectra.

\subsection{Observations}
\label{subs:observations}

We carried out the observations during the commissioning of the FMLO system on the Nobeyama 45-m telescope in early June 2016 and 2017 using the TZ front-end receiver and SAM45 backend spectrometer.
With both FMLO and PSW observations, we configured the A7 array of SAM45 in LSB ($\nu\msp{RF}=97.98097$~GHz).
We set the spectral channel spacing of SAM45 to 0.48828~MHz and the total bandwidth as 2000~MHz (4096~channels in total), which respectively correspond to 1.50~km/s and 6118~km/s in velocity at the observed frequency of 98~GHz.

\begin{figure*}[t]
  \centering
  \includegraphics[width=\linewidth]{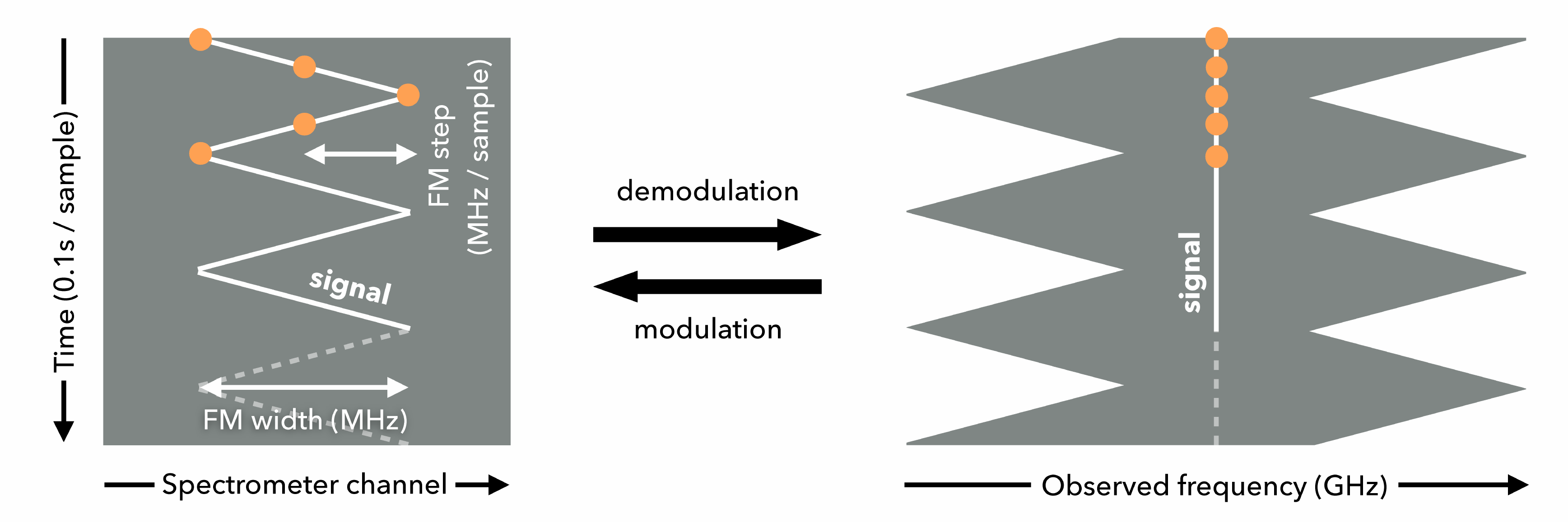}
  \caption{
    Schematic diagram of a zig-zag FM pattern.
    (left) A modulated timestream where the orange dots represent modulated signals fixed at the RF frequency.
    The FM width is the total modulation width over a timestream, and the FM step is an interval between successive time samples.
    (right) A demodulated timestream where signals are aligned to the RF frequency.
  }
  \label{fig:zigzag-fmp}
\end{figure*}

With the FMLO observations, we recorded the output timestream data of the on-point at a rate of 10~Hz by the SAM45 spectrometer.
At that time, we used a zig-zag-shaped function as the FM pattern, which has two free parameters, an FM width, and an FM step, as illustrated in figure~\ref{fig:zigzag-fmp}.
By definition, the total observed bandwidth is the sum of the total band width of the spectrometer and the FM width.
This means that a wider FM width results in a wider total observed bandwidth but fewer samples at the edge of the demodulated timestreams.
Thus, the sensitivity loss at the edge is greater than that at the center of the observed band.
On the other hand, a narrower FM width or shorter FM step may fail to estimate correlated and non-correlated components by PCA, when the frequency width of a target spectral line is wider than them, which might produce incorrect estimates of the spectral line.
We, therefore, choose these two parameters such that the FM width is wider than the FWHM of a spectral line and the FM step is as wide as possible within the FM width in order to optimize the conditions above.

\subsection{Data reduction}
\label{subs:data-reduction}

In the offline data reduction after an observation, we conducted 2-channel binning of the timestream data to reduce the total number of channels by half to 2048 to reduce the computation time.
We calibrated the absolute intensity of the data by the one-load chopper wheel method to derive atmospheric opacity-corrected antenna temperatures.
In the case of the PSW data, we subtracted a linear baseline from each spectrum to make a final product.

\begin{figure*}[t]
  \centering
  \includegraphics[width=\linewidth]{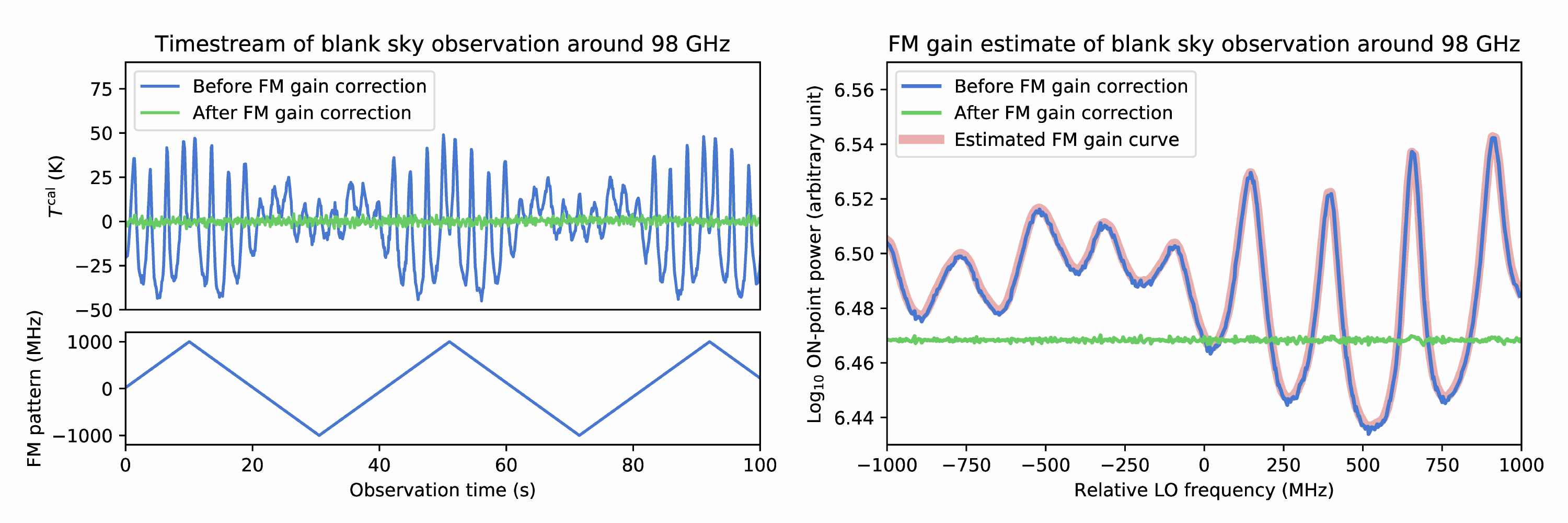}
  \caption{
    The demonstration of FM-dependent gain correction in a timestream in LSB around 98~GHz.
    The left panels show the antenna temperature of a modulated timestream, $\bT\msp{cal}$, at a channel of the band center and the corresponding FM pattern with observed time.
    The function of FM pattern in the antenna temperature without FM-dependent gain correction (blue line) shows a clear trend.
    The right panel shows the power of the on-point at the same channel in the relative LO frequency versus log power space, where relative LO frequency is expressed as $\nu\msp{LO}(n)-\nu\msp{LO,0}$.
    The resulting FM-dependent gain curve by a Savitzky--Golay smoothing filter (pink line) is overlaid on the raw power (blue line).
    Both panels show the antenna temperature and power after correcting the FM-dependent gain (green lines).
  }
  \label{fig:fmgain-correction}
\end{figure*}

In the case of the FMLO data, we found that the power of the on-point timestream changes as the frequency is modulated, which indicates that the gain between power and temperature is not constant during an observation period between chopper measurements (typically, $\sim$15--30~minutes).
We corrected for this FM-dependent gain by using the timestream data itself before the absolute intensity calibration:
we applied the Savitzky--Golay smoothing filter (\cite{Savitzky1964}; window length of 51; polynomial order of 3) to a timestream of an on-point assuming that the gain change is only a function of the frequency modulation.
Figure~\ref{fig:fmgain-correction} demonstrates that the FM-dependent gain curve of an observation is fitted in the $\nu\msp{LO}$--power space and is corrected in the calibrated timestream, $\bT\msp{cal}$.

We then conducted the data reduction procedure of the FMLO method as described in section~\ref{subs:data-reduction-procedure}.
We chose the number of principal components to model correlated components, $K$, such that the line free channels of each timestream spectrum were flat enough to estimate the noise-free product (typically, $K \simeq 5$ is used to conduct a $\sigma$-cutoff at $\theta\msb{cutoff}=5$).
We also chose the threshold value of convergence as $\varepsilon = 0.05$.
In the case of a mapping observation, the atmospheric condition may have changed as the elevation of an antenna was not constant during an observation, which suggests that the eigenvectors of correlated components, $\bP$, should be frequently estimated in a short period (several minutes).
We split a timestream into many time-chunks so that the time length of each chunk should be short enough ($N\sim600$; $\sim$10~minutes) and applied correlated component removal to them independently.

We note that there could exist pointing errors between the PSW and FMLO observations of the same target induced by wind loads, temperature variation, and time-dependent deformation of the telescope dish.
These errors cannot be corrected because we cannot observe them simultaneously with both methods, which may result in an intensity fluctuation in the spectral line emission between two observations (typically $\pm$10~\% in our commissioning).
In the following subsections, we therefore discuss the consistency of an FMLO observation with that of a PSW observation taking intensity fluctuation into consideration.

\subsection{Blank sky observation}
\label{subs:blank-sky-observation}

Before turning our attention to the astronomical sources, we observed a blank sky (i.e., an off-point) with the FMLO method where no astronomical spectral lines are expected to exist.
Such an observation can minimize the effect of astronomical signals and thus is suitable for the demonstration of, in particular, measuring noise levels and observation efficiency.
We used an FM pattern whose FM width was 2000~MHz and FM step was 10~MHz/sample.

\begin{figure*}[t]
  \centering
  \includegraphics[width=\linewidth]{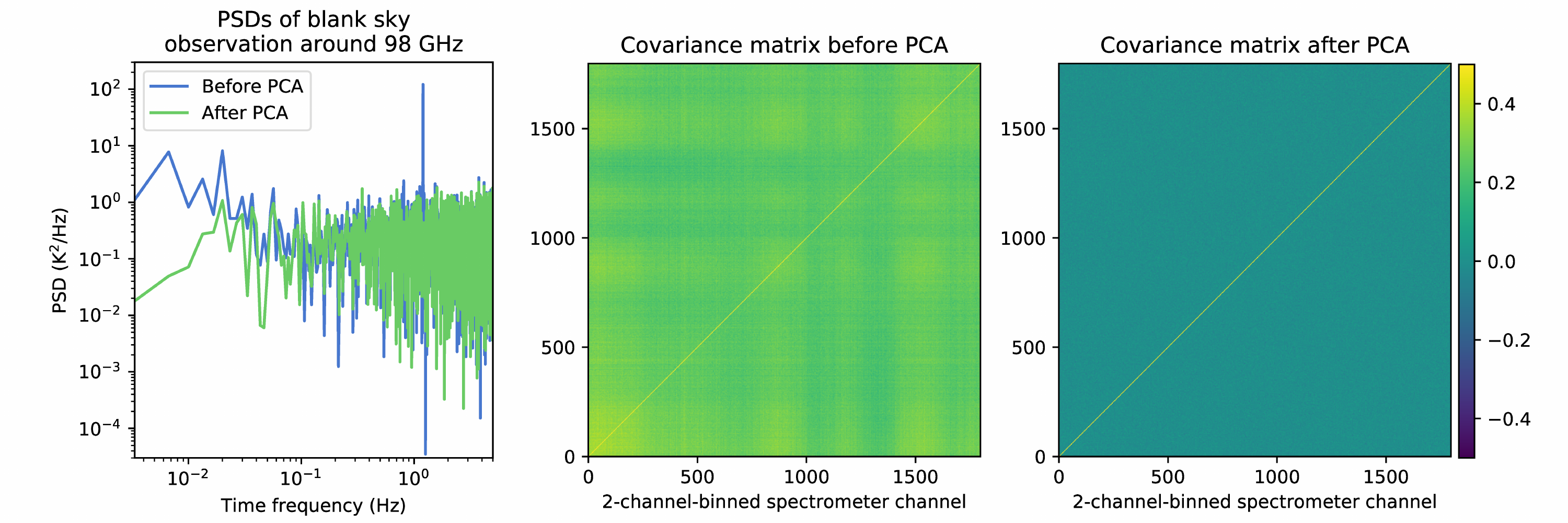}
  \caption{
    (left) The power spectrum densities (PSDs) of the 98~GHz channel before and after PCA cleaning.
    Note that a strong line-like feature seen at $\sim$1.2~Hz is attributed to a periodic baseline bobbing caused by variation of a mechanical chiller for a heterodyne receiver.
    (center) The covariance matrix created by $\bT\msp{cal}$ (before PCA cleaning).
    (right) The covariance matrix created by $\bT\msp{cln}$ (after PCA cleaning).
    Note that the timestreams are normalized so that diagonal values of the derived covariance matrix are unity.
  }
  \label{fig:blanksky-psd-cov}
\end{figure*}

We verify how correlated component removal reduces low frequency noises ($\lesssim 10$~Hz) in a cleaned timestream, $\bT\msp{cln}$, by measuring the power spectral densities (PSDs) and covariance matrices of a timestream before and after PCA cleaning.
Figure~\ref{fig:blanksky-psd-cov} shows the results:
As is seen in the covariance matrix before PCA cleaning, correlated components remain in the timestream, $\bT\msp{cal}$.
After correlated component removal, low frequency noises at $\lesssim$ 0.1~Hz decrease by 1--2 orders of magnitude.
A covariance matrix after PCA also shows no correlated components remaining compared to diagonal (auto correlation) values.

\begin{figure*}[t]
  \centering
  \includegraphics[width=\linewidth]{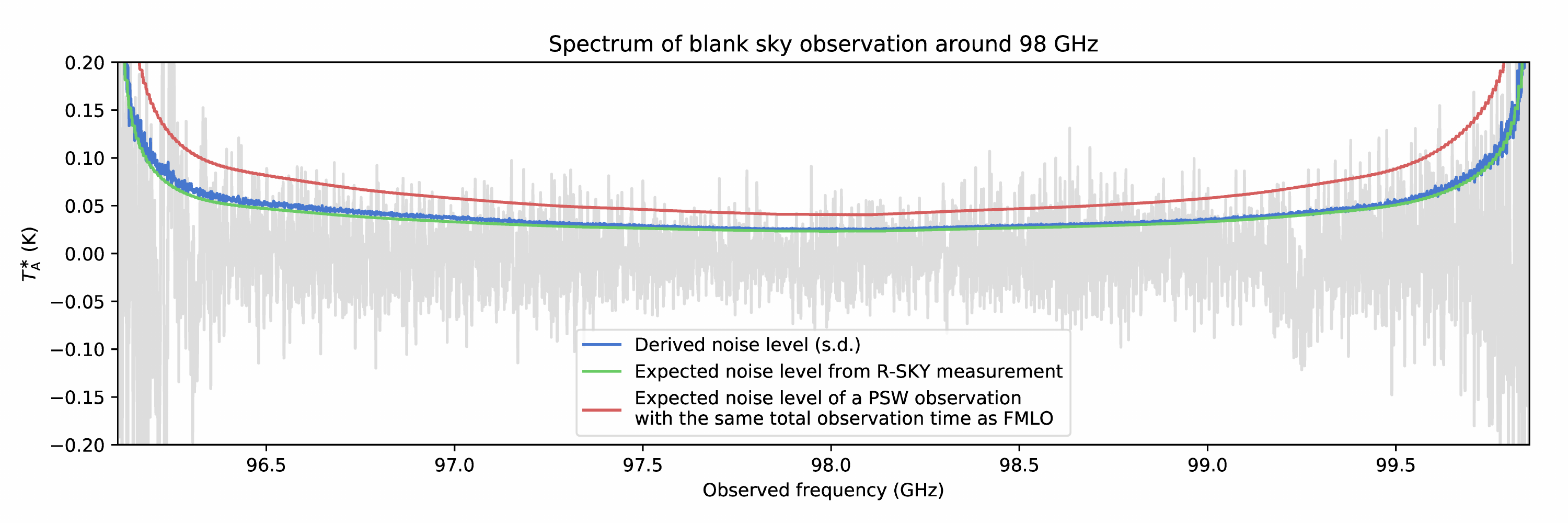}
  \caption{
    The final spectrum of a blank sky around 98~GHz (LSB) with atmospheric line emission subtracted (light blue line).
    We also plot (a) 1$\sigma$ noise level (standard deviation) derived from the timestream itself (green line), (b) 1$\sigma$ noise level expected to be achieved in R-SKY measurements (i.e., calculated from the equation~\ref{eqn:fmlo-sensitivity} with $\alpha=1$) (red line), and (c) 1$\sigma$ noise level expected to be achieved with a PSW observation of the same total observation time as the FMLO observation (purple line).
    We derive the factor of noise contribution from correlated component removal, $\alpha\msb{FMLO}\sim1.1$, over the observed band estimated by dividing (a) by (b).
    Note that the two spectral dents seen at 96.3~GHz and 99.25~GHz are caused by atmospheric ozone lines, the subtraction of which are discussed in section~\ref{subs:FMLO-method-in-more-generalized-cases}.
  }
  \label{fig:blanksky-spectrum}
\end{figure*}

Figure~\ref{fig:blanksky-spectrum} shows the final spectrum of a demodulated timestream, the 1$\sigma$ noise level evaluated from the timestream itself calculated by using the equation~\ref{eqn:fmlo-sensitivity} (with $\alpha\msb{FMLO}=1$), and the 1$\sigma$ noise level which is expected to be achieved with a PSW observation of the same observation time.
The noise level curve of a timestream is estimated by the bootstrap method by randomly changing the signs of samples of demodulated residual timestreams to resample the final spectra and derive the standard deviation.
As the number of samples ($\propto$ on-source time) of each frequency channel of the demodulated timestreams depends on the FM pattern, the noise level gets worse near the edge of the spectrum.
As a result, the factor of noise contribution from correlated component removal is achieved as $\alpha\msb{FMLO}\sim1.1$ over the observed band.
In other words, \emph{equivalent} noises from the off-point are $\Delta T\msb{off}=\sqrt{1.1^{2}-1^{2}}\,(T\msb{sys}/\sqrt{\Delta\nu\, t\msb{on}}) \sim 0.46T\msb{sys}/\sqrt{\Delta\nu\, t\msb{on}}$, which means that the accurate estimates of the in-situ baseline achieved with the FMLO method more than double.
We will discuss the actual value of $\alpha\msb{FMLO}$ in section~\ref{s:discussion}.

The achieved improvement in sensitivity of the FMLO method according to the equation~\ref{eqn:fmlo-sensitivity-improvement} is $\iota=1.74$, or the FMLO observation requires only $1/\iota^{2}=33$~\% of total observation time compared to that of the PSW to achieve the same sensitivity of the final spectrum.
In other words, we can \emph{equivalently} observe with a telescope whose system noise temperature is $(1-1/1.74)\sim43$~\% lower than the previous one.

In the actual observations, observation efficiencies of both FMLO and PSW methods are lower than the ideal ones.
For example, in the blank sky observations we used for the verification, $\eta\msb{obs}\msp{FMLO}$ and $\eta\msb{obs}\msp{PSW}$ were 0.69 and 0.42, respectively.
This is because, compared with typical scientific observations (on-source time of several hours), both observations are short (on-source time of 5~min) and the fraction of the overhead, such as the initial and final procedures of an observation, is large.
Using the values of actual observing efficiencies, however, we achieved an improvement of $\iota=1.65$, which is almost the same value as the ideal one.
We also note that the derived improvement, $\iota$, from this commissioning is the lower limit:
When we conduct a scientific observation, $\eta\msb{obs}\msp{PSW}$ is going to be much smaller because of the larger fraction of telescope slew time between the on- and off-points since the single observation time of each point should be short ($<10$~s, for example) for better subtraction between two points.

\subsection{Single-pointed observation}
\label{subs:single-pointed-observation}

We observed CS~$J=2-1$ (hereafter CS~(2--1); $\nu\msb{rest}=97.980953$~GHz; in LSB) of a carbon rich star, IRC~+10216, with an FM pattern whose FM width was 250~MHz and FM step was 80~MHz/sample.
The FM pattern fulfills the conditions above because the line width (full width at zero intensity; FWZI) of IRC~+10216 is expected to be $\sim40$~km/s (13~MHz) from past PSW observations (e.g., \cite{Cernicharo2010}).
The on-source time was $40\times \eta\msb{obs}\msp{FMLO}$~s (400~samples) and the achieved noise level of the FMLO observation per spectral channel, $\Delta T\msb{FMLO}$, was 0.046~K.
Our references were PSW observations carried out four times at about an hour before and after the FMLO observation with the same observation conditions as the FMLO observation.
Within each observation, we obtained 10~s of on and off-point observations four times to achieve the on-source time of 40~s.
The coordinates of the off-point were taken at 6-arcmin west of the on-point.
The achieved noise level of the PSW observation adjacent to the FMLO observation was $\Delta T\msb{PSW}=$0.057~K.

\begin{figure*}[t]
  \centering
  \includegraphics[width=\linewidth]{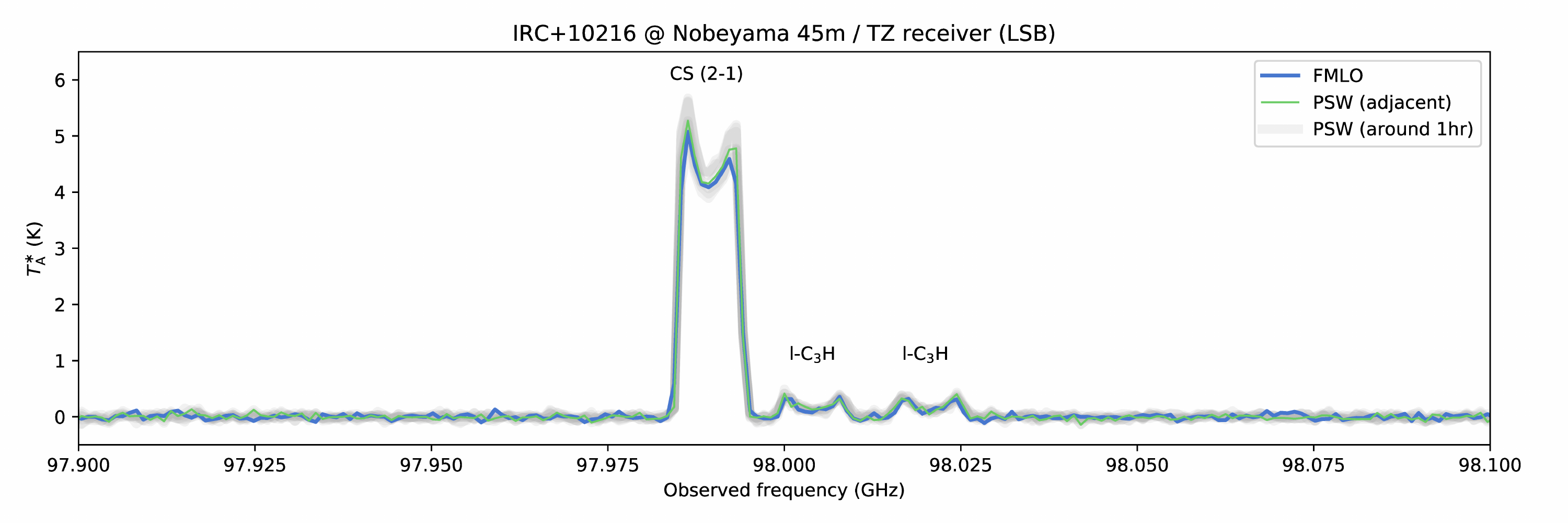}
  \caption{
    The obtained CS~(2--1) spectra of IRC~+10216 seen at 97.989~GHz observed with PSW (green line) and FMLO (blue one) methods at Nobeyama 45-m.
    We also plot various PSW spectra obtained around 60~min before and after the FMLO observation (gray lines) for monitoring typical pointing errors.
    Note that other two line features seen at 98.004 and 98.020~GHz are l-C$_{3}$H \citep{Mauersberger1989,Agundez2012}.
  }
  \label{fig:irc+10216-spectrum}
\end{figure*}

Figure~\ref{fig:irc+10216-spectrum} shows the resulting FMLO spectrum of CS~(2--1).
We also show the PSW spectra, which were observed adjacent to the FMLO observation.
These results show that the intensity can be easily changed within a hour beyond the noise level when we see a point-like source.
Based on comparisons of the FMLO and PSW spectra taken at different times, however, we can still confirm that both the intensity and line shape of the FMLO spectra are consistent with those of PSW, as the intensity of the FMLO spectrum is between the minimum and the maximum of those of the PSW.
If we demonstrate the FMLO method for deeper spectral observation (mK order of noise level) in a future commissioning, it would be necessary to confirm the consistency of the FMLO method with the ``time series'' PSW measurements.

Finally, we estimate $\alpha\msb{FMLO}$ from those observations using the following equation:
\begin{equation}
  \frac{\sqrt{2}}{\alpha\msb{FMLO}} = \frac{\Delta T\msb{PSW}}{\sqrt{\eta\msb{obs}\msp{FMLO}} \Delta T\msb{FMLO}},
\end{equation}
which yields $\alpha\msb{FMLO} = 1.10$ for the 45-m observations.
This is consistent with the $\alpha\msb{FMLO}$ measured in the blank sky observations.

\subsection{Mapping observation}
\label{subs:mapping-observation}

\subsubsection{Comparison}

We made a raster-scan mapping observation of CS~(2--1) toward the Orion KL $10 \times 10$~arcsec$^{2}$ region using the Nobeyama 45-m telescope.
We chose the FM pattern whose FM width is 120~MHz and FM step is 40~MHz/sample.
The FM pattern fulfills the conditions of the optimal FM pattern because the FWZI of Orion KL is expected to be $\sim40$~km/s (15~MHz) from past PSW observations (e.g., $^{12}$CO data cube of \cite{Shimajiri2011}).
Both conventional OTF and FMLO mapping observations were carried out with two raster-scan patterns;
x-scan (each scan is made along right ascension axis) and y-scan (declination axis).
The detailed parameters of the patterns are summarized in table~\ref{tab:scanpattern}.
The typical system noise temperatures, $T\msb{sys}$, during these observations were 230~K (LSB).
As will be further described, the on-source time per spatial grid was 2.66~s (a factor of 92~\% is included) and the achieved noise level of the FMLO observation per spectral channel was 0.18~K (LSB) after applying map making and basket-weaving methods \citep{Emerson1988}.
Before and after the two FMLO mapping observations, conventional OTF mapping observations were carried out (the x-scan mapping was before and the y-scan mapping was after them).
The typical $T\msb{sys}$ during these observations was 210~K (LSB).
The on-source time per spatial grid was 2.90~s, and the achieved noise level of the FMLO observation per spectral channel was 0.15~K (LSB) after applying map making and basket-weaving methods.
The coordinates of the off-point was 30--arcmin east of the center of the mapping region.

\begin{figure*}[t]
  \centering
  \includegraphics[width=\linewidth]{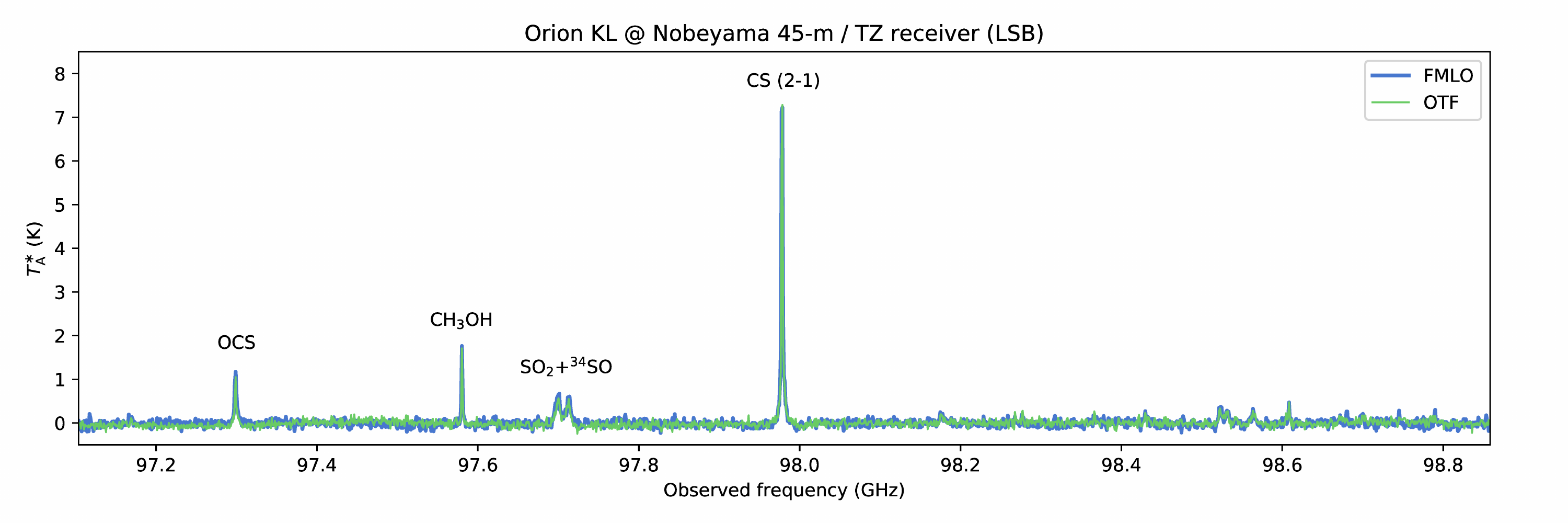}
  \caption{
    Mean spectra inside a 30–arcsec radius of Orion KL by the observations of OTF mapping (green line) and the FMLO mapping (blue line), respectively.
    Note that line features of sufficient signal-to-noise ratios are identified by the molecular line survey of \citet{Turner1989}.
  }
  \label{fig:orion-kl-spectrum}
\end{figure*}

\begin{figure*}[!h]
  \centering
  \includegraphics[width=\linewidth]{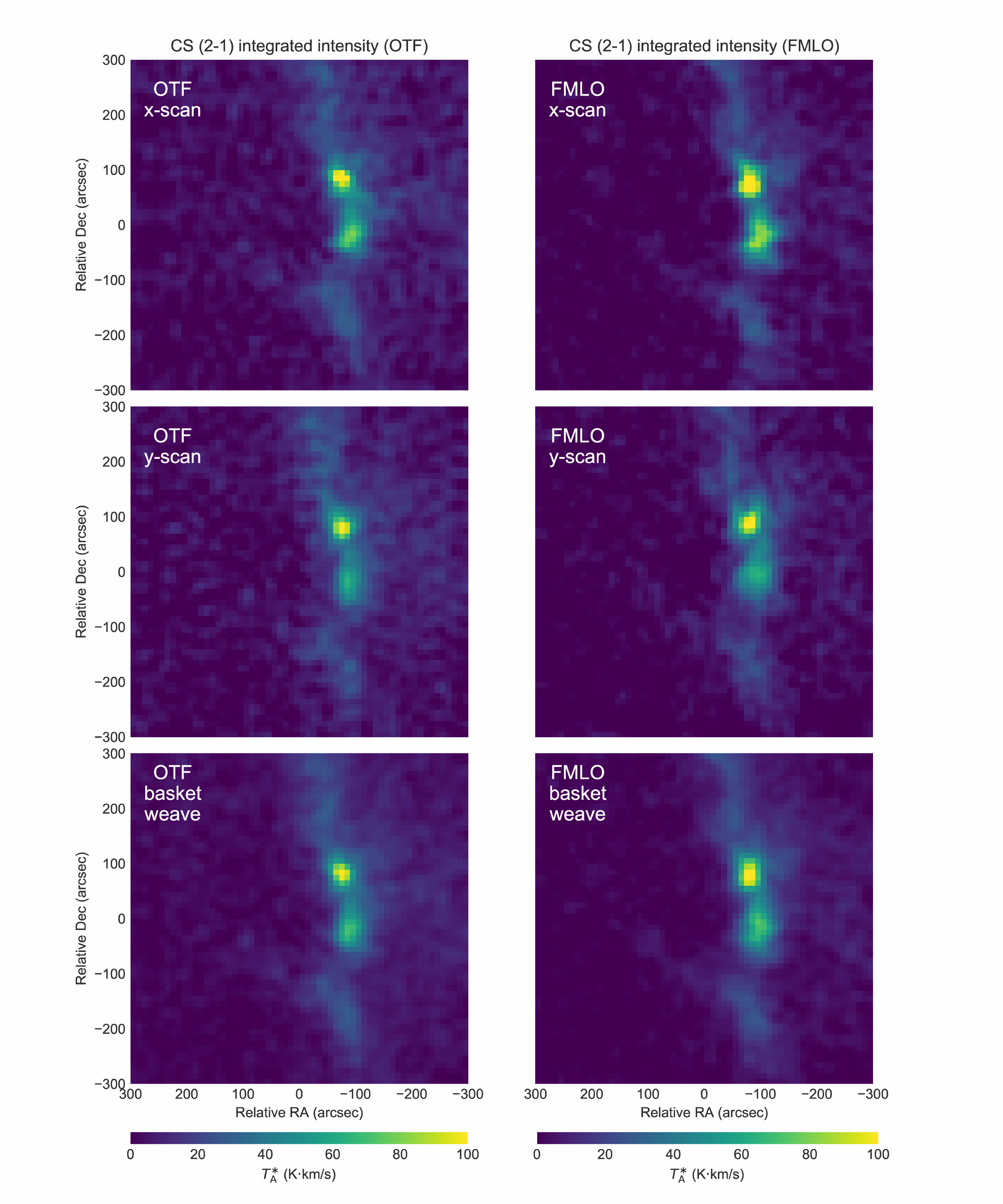}
  \vspace{3mm}
  \caption{
    The integrated intensity maps of CS~(2--1) of the $10\times10$~arcmin$^{2}$ Orion region.
    The upper four panels are maps of single scan directions with both the OTF and FMLO methods.
    The velocity range used for the integration is $v\msb{LSR}=[-16.25, +34.25]$~km/s, which contains both line and line-free velocity channels.
    The bottom two panels are maps with both OTF and FMLO methods after basket-weaving derived from maps of two different scan directions in order to minimize the scanning effect.
    The velocity range used for the integration is $v\msb{LSR}=[-6.25, +24.25]$~km/s.
  }
  \label{fig:orion-kl-maps}
\end{figure*}

After applying the basket-weaving method, we obtained a final FMLO map (3D cube) that is expected to be consistent with that of the OTF method (and also $T\msb{A}^{\ast}$).
Figure~\ref{fig:orion-kl-spectrum} shows the spectra obtained with the OTF and FMLO methods by averaging the spectra inside a 30--arcsec radius of Orion KL.
A comparison of the spectra obtained by the OTF and FMLO methods reveals that the obtained FMLO spectrum is almost consistent with that of the OTF.
Figure~\ref{fig:orion-kl-maps} shows the integrated intensity maps of CS~(2--1) created from the x-scan, y-scan, and basket-weaved 3D cubes.
Comparisons between the OTF and FMLO maps of each row of figure~\ref{fig:orion-kl-maps} reveal that the overall spatial distribution and intensity of the FMLO maps are almost consistent with those of the OTF maps.
Moreover, we demonstrate that the scanning effect (one of ``correlated'' components) seems to be removed in each raster scan of a single direction, while the large scanning effect remains in the the x- and y-scan of OTF maps along with their scan patterns before the basket-weaving method is used.
After basket-weaving, both strong and weak structures seem to be consistent with each other.
We note that the overall structures of four maps seem to be slightly shifted from each other, which indicates that there exist pointing errors between them.
We estimate the maximum pointing error by comparing the pixel coordinates of the maximum intensity values; as result, at most 10~arcsec (1 pixel) of the distances can be shifted.
If we observe a point-like source in a 98~GHz (FWHM beam size of 17~arcsec, assuming Gaussian shape) by pointing 10~arcsec away from the source, the intensity will be about 30~\% of the intrinsic value ($\sim3$ times change).
In the following analysis, we thus use the $3\sigma$ noise level as the standard deviation value.

\begin{figure}[t]
  \centering
  \includegraphics[width=\linewidth]{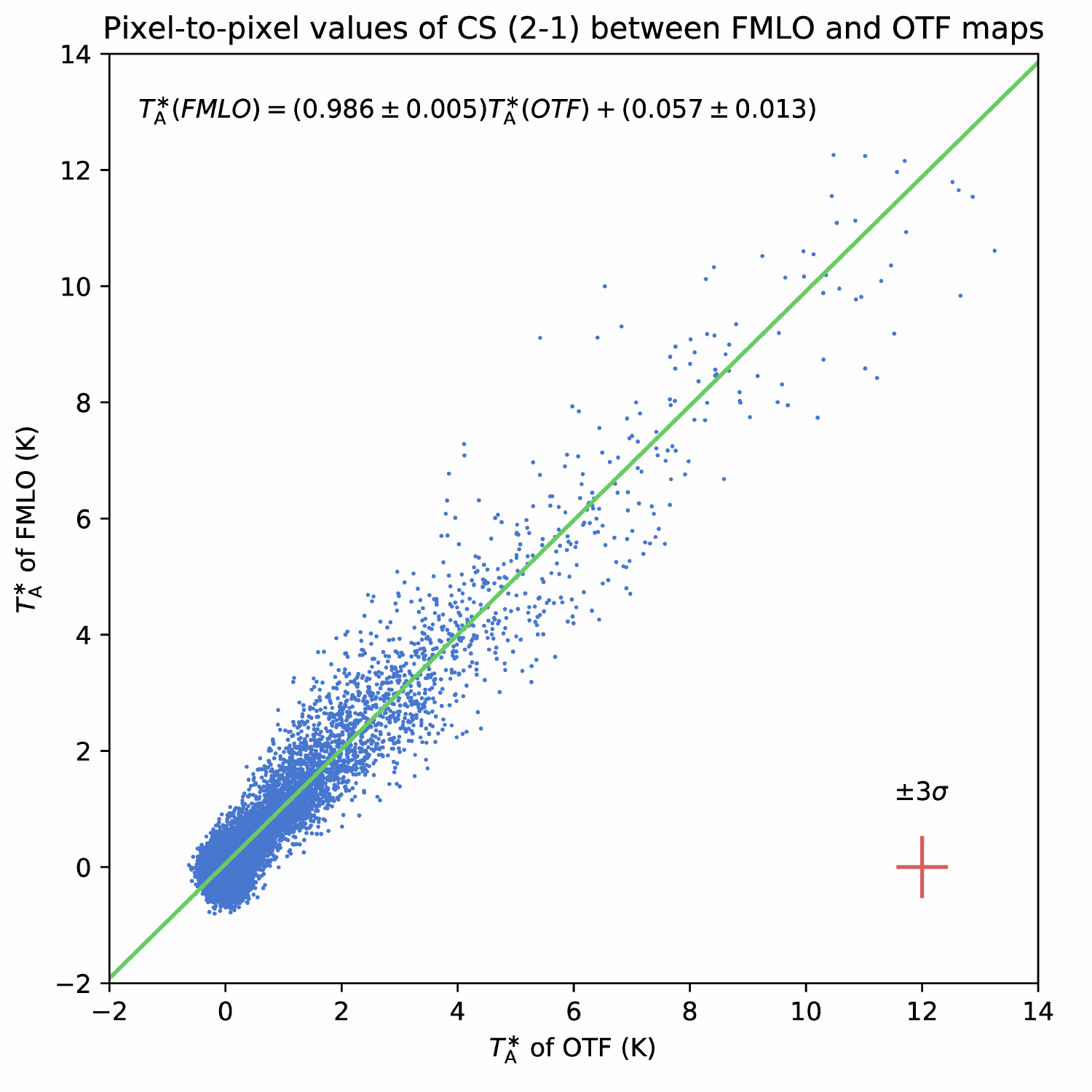}
  \caption{
    The pixel-to-pixel correlation plot between the OTF and FMLO mappings of CS (2-1) in LSB.
    Pixels are $d\times d\times \Delta\nu$ elements of a 3D cube with $d=10$~arcsec, $\Delta\nu=0.977$~GHz, and we plot the pixels with a velocity range of $v\msb{LSR} = [-6.25, +24.25]$~km/s around CS~(2--1).
    Linear fit of ($y=ax+b$) is conducted by an orthogonal distance regression with x and y errors of a $3\sigma$ noise level derived from the 3D cube (see also table~\ref{tab:noiselevel-maps}).
    The results of the fit are displayed at the top of panel.
  }
  \label{fig:orion-kl-p2p}
\end{figure}

To confirm the consistency between OTF and FMLO mapping observations within several uncertainties (noise level, pointing errors, and intensity calibration), we create a pixel-to-pixel correlation plot between them.
This approach is used in \citet{Sawada2008} to confirm the consistency between the OTF and PSW methods:
We aim to confirm the consistency between the FMLO and OTF mappings within the accuracy of a relative intensity calibration of 5\%, which is required for an intensity reproducibility of the standard source.
Figure~\ref{fig:orion-kl-p2p} shows the pixel-to-pixel scatter plot and a line fit of $T\msb{A}^{\ast} (\mr{FMLO})= a\, T\msb{A}^{\ast} (\mr{OTF}) + b$ to data points.
The frequency range of pixels selected is the same as the one used for creating integrated intensity maps of basket-weaved data ($-6.25<v\msb{LSR}<24.25$~km/s).
The $3\sigma$ noise levels of OTF and FMLO are used for calculating uncertainties of ($a, b$) in line fittings.
The results reveal that the correlation coefficient, $a$, is $0.986\pm0.005$, which suggests that the OTF and FMLO maps are consistent within 1.4\%.

\subsubsection{Achieved improvement}

Finally, we demonstrate the sensitivity improvements of the FMLO mapping compared to that of OTF in the same manner described in section~\ref{subs:observation-efficiency-and-sensitivity-improvement}.
For the FMLO mapping observation, we use $\alpha\msb{FMLO}\simeq1.1$, which is estimated by comparing the achieved noise level and that calculated from $T\msb{sys}$ in section~\ref{subs:blank-sky-observation}.
For the OTF mapping observation, $\alpha\msb{OTF}$ is expressed as
\begin{equation}
  \alpha\msb{OTF}
  = \sqrt{1+\frac{t\msp{on}\msb{cell}}{t\msp{off}\msb{cell}}},
\end{equation}
where $t\msp{on}\msb{cell}$ and $t\msp{off}\msb{cell}$ are on- and off-source integration times per a spatial grid cell introduced by \citet{Sawada2008}:
\begin{eqnarray}
  t\msp{on}\msb{cell}
    &=& \frac{\eta d^{2}}{l_{1}l_{2}}\,t\msp{on}\msb{total},
  \label{eqn:t_on_cell}\\
  t\msp{off}\msb{cell}
    &\simeq& \frac{d}{\Delta l}\,t\msb{off},
  \label{eqn:t_off_cell}
\end{eqnarray}
where $t\msp{on}\msb{total}$ is the total on-source integration time of a mapping observation, $t\msb{off}$ is an single integration time of an off-point, and $\eta$ (not an observation efficiency) is a factor determined by the extent of the used GCF.
$t\msp{on}\msb{total}$ is a product of the number of scans, $N\msb{row}\, (=l_{2}/\Delta l + 1)$, and an observed time of a scan, $t\msb{scan}$.
The value of $\eta$ for a Bessel--Guass GCF with default parameters is 4.3 \citep{Sawada2008}.
The values of $t\msp{on}\msb{cell}$ and $t\msp{off}\msb{cell}$ are summarized in table~\ref{tab:scanpattern}, which yields $\alpha\msb{OTF}=1.04$.
From table~\ref{tab:noiselevel-maps}, we can confirm that the calculated noise levels per spatial grid per frequency channel are almost consistent with those of actual values derived from 3D cubes themselves.
We then derive the observation efficiencies, $\eta\msb{obs}\msp{map}\, (=t\msp{on}\msb{total}/t\msp{obs}\msb{total})$, of the OTF and FMLO maps.
Unlike for spectral line observations (single-pointed), we need to take into account several overheads, such as the preliminary antenna movement necessary for initializing and finalizing a scan measurement.
From \citet{Sawada2008}, the total on-source time, $t\msp{on}\msb{total}$, and total observation time, $t\msp{obs}\msb{total}$, can be expressed as follows:
\begin{eqnarray}
  t\msp{on}\msb{total}
    &=& N\msb{row}t\msb{scan}\\
  t\msp{obs}\msb{total}
    &=& N\msb{row}\brp{
      t\msb{scan} + t\msb{OH} +
      \frac{t\msb{off}}{N\msb{scan}\msp{seq}}
    }f\msb{cal},
\end{eqnarray}
where $t\msb{OH}$ is the overhead time per scan, $N\msb{scan}\msp{seq}$ is the number of scans taken between off-point measurements, and $f\msb{cal}$ is a dimensionless factor that represents the overhead of the chopper wheel calibration.
$t\msb{OH}$ can be expressed as the sum of several overhead terms:
\begin{equation}
  t\msb{OH}
    = \frac{2t\msp{off}\msb{tran}}{N\msb{scan}\msp{seq}}
    + t\msb{app}
    + \frac{N\msb{scan}\msp{seq}-1}{N\msb{scan}\msp{seq}}\,t\msb{tran},
\end{equation}
where the first, second, and third terms correspond to the time the antenna slew between the on and off-points, antenna approach time for initializing a scan measurement, and transition time for finalizing a scan, respectively.
The values of these parameters are summarized in table~\ref{tab:scanpattern}.
These yield $\eta\msb{obs}\msp{map}=0.50$ with the FMLO map and 0.39 with the OTF map.
Together with the noise contribution, $\alpha$, we achieve the sensitivity improvement of $\iota=1.07$ and that for the observation time of $\iota^{2}=1.15$ from these calculations.
Moreover, these values from actual $t\msp{on}\msb{total}$ and $t\msp{obs}\msb{total}$ are $\iota\msb{real}=1.11$ and $\iota\msb{real}^{2}=1.23$, which demonstrates that an FMLO mapping is 23\% more efficient than that of OTF with regard to unit noise level, although an OTF mapping observation is more efficient than a PSW observation.
We note that the mapping region we used for the commissioning (100~arcsec$^{2}$) is orders of magnitude smaller than the known typical mapping surveys by an order of magnitude (\citet{Shimajiri2014} conducted an observation of 1440~arcsec$^{2}$, for example), which results in an improvement that is smaller than that of a single-pointed observation ($\iota\sim1.7$).
We will discuss the expected improvement in more realistic, larger mapping observations in section~\ref{s:discussion}.

\begin{table}[t]
  \centering
  \tbl{Standard deviation noise levels per pixel of the OTF and FMLO final maps.}{
    \begin{tabular*}{\linewidth}{@{\extracolsep{\fill}}|r||r|r|}
      \hline
      & OTF & FMLO\\
      \hline\hline
      Expected $\Delta T\msb{A}^{\ast}$ (K) & 0.13 & 0.16\\
      Derived $\Delta T\msb{A}^{\ast}$ (K) & 0.15 & 0.18\\
      \hline
    \end{tabular*}
  }
  \label{tab:noiselevel-maps}
  \begin{tabnote}
    The expected values are derived from equation~\ref{eqn:on-off-sensitivity} using $\alpha$, $t\msb{cell}\msp{on}$ and $t\msb{cell}\msp{off}$ of each method ($\alpha\msb{OTF}=1.04$ and $\alpha\msb{FMLO}=1.1$, respectively).
    The values of $t\msb{cell}\msp{on}$ and $t\msb{cell}\msp{off}$ are summarized in table~\ref{tab:scanpattern}.
  \end{tabnote}
\end{table}

\begin{table*}[t]
  \centering
  \tbl{Observational and data reduction parameters of OTF and FMLO observations.}{
    \begin{tabular*}{\textwidth}{@{\extracolsep{\fill}}|r||r|r||r|r|}
      \hline
      & OTF ($10 \times 10$~arcmin$^{2}$) & FMLO ($10 \times 10$~arcmin$^{2}$) & OTF ($1 \times 1$~deg$^{2}$) & FMLO ($1 \times 1$~deg$^{2}$)\\
      \hline\hline
      $l_{1}$ (arcsec)                & 600 & 600 & 1200 & 3600\\
      $l_{2}$ (arcsec)                & 600 & 600 & 1200 & 3600\\
      $\Delta l$ (arcsec)             & 6 & 6 & 6 & 6\\
      $t\msb{scan}$ (second)          & 12 & 12 & 24 & 72\\
      $t\msb{off}$ (second)           & 12 & \textbf{0} & 24 & \textbf{0}\\
      $t\msb{tran}$ (second)          & 5 & 5 & 5 & 5\\
      $t\msp{off}\msb{tran}$ (second) & 10 & \textbf{0} & 10 & \textbf{0}\\
      $t\msb{app}$ (second)           & 5 & 5 & 5 & 5\\
      $N\msb{scan}\msp{seq}$          & 3 & \textbf{101} & 1 & \textbf{601}\\
      $f\msb{cal}$                    & $\simeq1$ & $\simeq1$ & $\simeq1$ & $\simeq1$\\
      \hline
      $d$ (arcsec)                    & 10 & 10 & 10 & 10\\
      $\eta$                          & 4.3 & 4.3 & 4.3 & 4.3\\
      \hline
      $t\msb{OH}$ (second)            & 15.0 & \textbf{10.0} & 25.0 & \textbf{10.0}\\
      $t\msp{on}\msb{cell}$ (second)  & \textbf{1.45} & 1.33 & \textbf{1.44} & 1.32\\
      $t\msp{off}\msb{cell}$ (second) & 20.0 & \textbf{0} & 40.0 & \textbf{0}\\
      $t\msp{on}\msb{total}$ (minute)  & \textbf{20.2} & 18.6 & \textbf{724} & 664\\
      $t\msp{obs}\msb{total}$ (minute) & 52.2 (58) & \textbf{37.0} (39) & 2201 & \textbf{821}\\
      $\eta\msb{obs}\msp{map}$         & 0.39 (0.35) & \textbf{0.50} (0.48) & 0.33 & \textbf{0.81}\\
      \hline
    \end{tabular*}
  }
  \label{tab:scanpattern}
  \begin{tabnote}
    Left two columns show the actual values used mapping observations toward the $10 \times 10$~arcmin$^{2}$ Orion-KL region.
    On the other hand, right two columns show the supposed values if we conduct $1 \times 1$~deg$^{2}$ mapping observations using OTF and the FMLO method, respectively.
    The three groups of rows mean as follows:
    (top) The observational parameters.
    If the values are different between OTF and FMLO, the better value is displayed with a bold symbol.
    (middle) The parameters of map making after obtaining mapping timestream data.
    $\eta$ (not an observation efficiency) is a factor determined by the extent of the used GCF ($\eta=4.3$ for a Bessel–-Gauss GCF with default parameters).
    (bottom) The derived values used for calculating observation efficiency.
    The values without parentheses are estimated values and those with parentheses are the actual values from timestream data and the observing logs.
  \end{tabnote}
\end{table*}

\section{Discussion}
\label{s:discussion}

\subsection{Advantages and limitations of the FMLO method}
\label{subs:advantages-and-limitations-of-the-fmlo-method}

\subsubsection{In-situ estimation of off-point without measurements}

We confirm that removing correlated noise is applicable to the (sub-)millimeter spectroscopy; this is one of the most important results of the study.
This suggests that the baseline spectrum of an off-point can be variable within a typical switching interval ($\sim5-10$~s) in the PSW method and obtaining time-series data more than $\gtrsim1$~Hz is necessary for the application.

As the FMLO method does not need to obtain any off-point measurements, it achieves remarkable sensitivity improvements of both spectral and mapping observations of the FMLO methods ($\iota\simeq1.7$ and 1.1, respectively) per unit observation time and noise levels.
We also confirm that the intensity and line shape of an astronomical spectral line are not affected by a correlated component removal by PCA if we choose an optimal FM pattern.
The approach of in-situ spectral baseline subtraction is therefore promising to remove the correlated noises from the atmosphere and obtain spectra with ideal noise levels.

Moreover, the FMLO method is an effective way for single-dish mapping observations:
an FMLO observation does not suffer from emission-line contamination of an ``off-point'', which is sometimes the case for abundant molecules such as CO.
As partially demonstrated in figure~\ref{fig:orion-kl-maps}, the FMLO method eliminates a ``scanning effect'' in a mapping observation only with a single scan pattern, which may also improve the sensitivity.

\subsubsection{Mitigation of instrumental noises and IF interference}

Another advantage of a correlated component removal is that it enables to detect spectral features appeared at a fixed IF frequency or over the entire observed waveband.
As demonstrated in figure~\ref{fig:blanksky-psd-cov}, it effectively subtracts periodic baseline bobbing caused by vibration of a mechanical chiller for a heterodyne receiver.
It would be also effective to mitigate a spurious signal at a certain spectrometer channel and artifitial interference by wireless communication in IF band.

\subsubsection{Continuum and broader line observations}

\begin{figure*}[t]
  \centering
  \includegraphics[width=\textwidth]{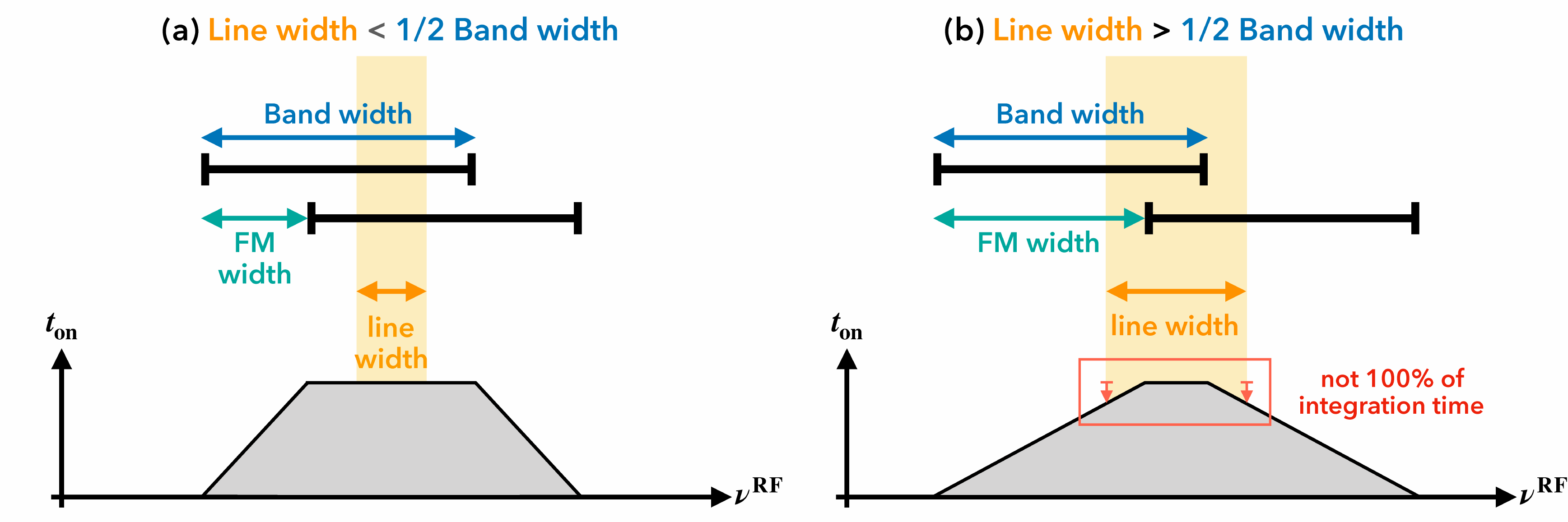}
  \caption{
    Schematic diagram of the requirements of the spectrometer's band width for spectral line observation.
    The top black bars show the band width of a spectrometer with frequency modulation.
    The bottom graphs show total on-source time achieved with a zig-zag FM pattern as a function of the observed RF frequency.
    (a) the case where the line width $<$ 1/2 band width: all the samples of a timestream fully cover the line width (FWHM) and sensitivity loss does not occur.
    (b) the case where the line width $>$ 1/2 band width: some samples of a timestream do not cover the line width (FWHM), and sensitivity loss occurs at the edges of the line.
  }
  \label{fig:spectrometer-requirements}
\end{figure*}

The optimal FM pattern for an FMLO observation depends on the line width (FWHM) of a target.
This, however, suggests that we need to know the intrinsic line width of a target by any means.
Although we can always choose and create an FM pattern of the largest FM width and step as the optimal one, we should carefully consider the case where the signal of an emission line occupies a large amount of the total band width of a spectrometer.
As an example, consider the spectral observation of a $\Delta v=300$~km/s CO line (assuming an extragalactic source). This is equivalent to $\sim0.1$~GHz for an observation of CO~(1--0), and it is narrow enough ($\sim6$~\%) for the total band width of the SAM45 spectrometer (\citet{Kamazaki2012}; 2000~MHz).
On the other hand, that is equivalent to $\sim0.3$~GHz for an observation of CO~(3--2), and it occupies $\sim70$~\% of the total band width of the MAC spectrometer (\cite{Sorai2000}; 512~MHz), which causes a little sensitivity loss at the edges of the line (see figure~\ref{fig:spectrometer-requirements}).
In order to eliminate such loss, the total band width of a spectrometer should be at least twice as wide as the line width.
As a consequence, obtaining continuum emission with an FMLO observation (corresponding line width $\gg$ observing bandwidth) is challenging.
With a multi-pixel heterodyne receiver such as FOREST \citep{Minamidani2016} and HERA \citep{Schuster2004}, however, it may be possible to obtain the continuum emission because not only are the astronomical signals modulated in the frequency domain, but also a spatial axis like (sub-)millimeter continuum camera is used.

We note that such sensitivity loss may not affect the band center much because it is only proportional to a square root of the total on-source time even though on-source time itself drops linearly if we use a zig-zag FM pattern.
We also note that this is not the case with a spectral line survey, which obtains a frequency range of several GHz over the instantaneous band width of a spectrometer.
In this case, we can set an FM width to be wider than half of the band width, which may be an efficient way to conduct the survey compared to conducting PSW observations several times by changing the center frequencies.

\subsection{FMLO method in more generalized cases}
\label{subs:FMLO-method-in-more-generalized-cases}

\subsubsection{Modeling atmospheric line emission}

In section~\ref{subs:data-reduction-procedure}, we describe the algorithm of data reduction for an observation where atmospheric line emission does not exit.
In more general cases where the atmospheric line emission contaminates the spectrum, the algorithm is naturally extended to estimate $\bT\msp{atm(,i)}$ by PCA.
The order of steps to estimate the components is as follows: $\bT\msp{cor} \rightarrow \bT\msp{atm} \rightarrow \bT\msp{ast} \rightarrow \bT\msp{atm,i} \rightarrow \bT\msp{ast,i}$.
This, however, also requires the PCA method to be extended so that the spectral channels with the atmospheric line emission can be de-weighted as they have a much broader line shape (FWZI $\gtrsim$ 1000 km/s) than the astronomical emission:
Such emission is considered \emph{correlated} components unless properly handled.
In the correlated component removal in more generalized cases, we introduce the weighted PCA by expectation maximization (EM) algorithm (EMPCA; \citet{Bailey2012}), which enables us to minimize the effect of strong line emission by de-weighting elements contaminated by such emission in a modulated timestream when estimating $\bP$ and $\bC$.
\citet{Bailey2012} presents the EMPCA for noisy data and/or data that is missing some points.
Noisiness and/or incomplete data are expressed as a weight matrix, $\bW$, which has the same dimension as the data matrix.
The classical PCA minimizes the following quantity of $\chi^{2}$:
\begin{equation}
  \chi^{2} = |\bX-\bP \,@\, \bC|_{F}^{2},
\end{equation}
while the EMPCA tries to minimize $\chi^{2}$ containing $\bW$:
\begin{equation}
  \chi^{2} = |\bW_{dn}(\bX-\bP \,@\, \bC)|_{F}^{2}.
\end{equation}
We will present an implementation of EMPCA and demonstrate the modeling of the atmospheric line emission in the next study.

\subsubsection{FM-dependent gain correction}

The conventional non-FM position switching method assumes that the gain, $\bG$ (also known as bandpass), does not change between the on-point and hot load measurements (i.e., $\bG\msp{on}\simeq\bG\msp{load}$).
In the FMLO method, however, the timestream of the on-point gain is modulated and thus has dependence on observed frequencies as we mention in section~\ref{subs:data-reduction}.
Before the absolute intensity calibration is performed, it is necessary to estimate the frequency-modulation-dependent gain, $\bG\msp{FM}$ and to separate it from $\bG\msp{on}$ to obtain the FM-independent bandpass.
In order to correct for the FM-dependent gain, we have two strategies:
\begin{enumerate}
  \item
    To obtain a timestream of the on-point with the FMLO method and a non-FM spectrum of the hot load (the method we adopt in this paper).
    FM-dependent gain is then estimated from the on-point timestream itself by smoothing the FM-dependent gain curve (as an analogy of self calibration in an interferometric observation).
  \item
    To obtain the timestreams of both the on-point and hot load.
    The FM-dependent gain is then estimated from the hot load measurements.
\end{enumerate}
In FMLO observations with the TZ receiver, the period of the gain curve is $\sim250$~MHz, which is comparable to the typical line width of extragalaxies (FWHM $\sim500$~km/s).
If we observe such targets, the latter method would be essential to distinguish the FM-dependent gain from the signal.
We will discuss both strategies using the observed FMLO data with the FOREST receiver in the next study.

\subsection{Improvement of sensitivity and efficiency}
\label{subs:improvement-of-sensitivity-and-efficiency}

\subsubsection{Effect of correlated component removal}

In section~\ref{subs:blank-sky-observation}, we demonstrate that PCA properly estimates correlated components and a cleaned timestream can be obtained after such components are subtracted.
There exists, however, an important issue regarding the contribution of noise from the correlated components themselves, which is expressed as the factor $\alpha\msb{FMLO}\simeq1.1$.
Although it is smaller than the factor of position switching ($\alpha\msb{PSW}=\sqrt{2}$), it would be better to minimize such a contribution (i.e., $\alpha\msb{FMLO}\rightarrow 1$).
One possible solution is smoothing the correlated components:
\citet{Bailey2012} discusses the possibility of ``smoothed EMPCA,''i.e., smoothing the basis vectors at each estimation step, and concludes that, compared to the smoothing of noisy eigenvectors, it will result in optimal smooth eigenvectors.
This, however, requires that the length scale of the intrinsic correlated components be larger than that of the noise, i.e., the correlated component should have a regular rather than a random spectral shape during an observation.
For such an approach, we should obtain data from different observation seasons to monitor the robustness of the modes of correlated components.
We may also derive an optimal window length for smoothing using the strategy of the smoothed bandpass calibration \citep{Yamaki2012}.

\subsubsection{Optimization of the modulation frequency}

We use the modulation frequency ($=$ dump rate of a spectrometer) of 10~Hz throughout the study.
While it is determined by an estimate of the time-scale of sky variation, the PSD plot in figure~\ref{fig:blanksky-psd-cov} suggests that correlated noises dominate in lower frequency than 0.1~Hz in the case of the Nobeyama 45-m telescope.
As the dwell time of a frequency sweep is fixed (section~\ref{subs:hardware-implementation}), the observation effeciency of FMLO would improve with a lower dump rate (for example, $\eta\msb{obs}\msp{FMLO}$ is 0.984 if a dump rate is 2~Hz, which results in $\sim$7\% improvement).
On the other hand, as also shown in the PSD plot, the existance of periodic vibration higher than 1~Hz should be also be considered, otherwise it may worsen the sensitivity.
The choice of the modulation frequency is therefore a trade-off.
The optimization of it for a telescope will be discussed in the next study.

\subsubsection{Application to large mapping observations}

In section~\ref{subs:mapping-observation}, we demonstrate that the sensitivity improvement of the FMLO mapping is $\iota\sim1.1$ (10~\% of improvement) compared to that of the OTF, which seems to be quite small in the case of a single-pointed observation of the FMLO method ($\iota\sim1.7$).
This is because, while the on-point observation of the FMLO is 20--30~\% more efficient, the noise contribution from the off-point, $\alpha\msb{FMLO}$, is 5\% worse than that from the OTF.
If $\alpha\msb{FMLO}$ does not increase, $\alpha\msb{OTF}/\alpha\msb{FMLO}$ worsens (less than unity) for a wider mapping area such as several square-degree surveys.
It is, however, expected that we will still obtain a better $\iota$ for a wider mapping area because there exists an upper limit on the scan length, $l_{1}$, of an OTF mapping observation owing to the upper limit of the observed time per scan, $t\msb{scan}$, while an FMLO observation has no such limit.
According to the ``OTF Observations with the 45-m Telescope'' of the Nobeyama 45-m website\footnote{\url{https://www.nro.nao.ac.jp/~nro45mrt/html/obs/otf/index_en.html}}, $t\msb{scan}$ should be 10--30~s and an off-point observation should be performed every 10--30~s for a mapping observation of the Nobeyama 45-m because (1) a larger time interval between off-points causes a baseline wiggle similar to that observed with the PSW method and (2) a longer $t\msb{scan}$ (i.e., longer observation time of an entire map) no longer guarantees the uniformity of a map.

Furthermore, there exists a lower limit for the scan speed, $v\msb{scan}$, where the spatial sampling interval should be 1/3--1/4 of the beam size (HPBW).
For a 115~GHz observation of the CO~(1--0) line (a beam size of 15~arcsec) using the Nobeyama 45-m, for example, $v\msb{scan}$ should be 50--60~arcsec/s, which yields an upper limit of $l_{1}=$ 1000--1200~arcsec.
If we  conduct a $1\times1$~degree$^{2}$ mapping observation using both the OTF and FMLO methods, for the OTF method, it is necessary to split the mapping area into nine different $20\times20$~arcmin$^{2}$ subregions.
This yields an observation efficiency of $\eta\msb{obs}\msp{map}=0.31$, while the FMLO achieves $\eta\msb{obs}\msp{map}=0.81$, a much higher value if we use the parameters described in table~\ref{tab:scanpattern}.
Together with the noise contribution factor, the sensitivity improvement is $\iota=1.45$ (45~\% of improvement) and the efficiency improvement per unit noise level is $\iota^{2}=2.10$ (110~\% of improvement).
This is because we assume that the FMLO mapping can break the upper limit of $t\msb{scan}$ and sweep a scan length of 1~deg at a time ($t\msb{scan}=72$~s).
We note that the derived $\iota$ is a lower limit:
In actual observations, we expect that the baseline wiggles and/or scanning effects are subtracted by correlated component removal, which will result in a much higher $\iota$ and thus guarantees the uniformity of a map with even a longer scan length.

\subsection{Computation cost for a data reduction}
\label{subs:computation-cost-for-a-data-reduction}

In a data reduction of the FMLO method, estimating correlated components by PCA takes the large amount of time.
A standard PCA requires a computation cost of $\mc{O}(\mr{min}(N^{3}, D^{3}))$, where $\mc{O}$ is big O notation to express the order of a fucntion.
If one needs to obtain only the first $K$ correlated components by singular value decomposition (SVD), the cost would be reduced to $\mc{O}(\mr{min}(ND^{2}, N^{2}D))$.
The typical size of a timestream in the study is $N=600, D=2048$.
For the first five correlated components ($K=10$), an estimate takes $\sim$220~ms (using Mac Pro Late 2013 with 3.5~GHz 6-core Intel Xeon E5).
Including any overheads such as loading data and other data reduction steps, the total reduction time of the single-pointed observation of IRC+10216 (section~\ref{subs:single-pointed-observation}) was $\sim$20~s with 16 iterations.
The total reduction time of the mapping observation of Orion KL (section~\ref{subs:mapping-observation}) is much longer than that of IRC+10216; the data size is thirty times larger ($N\sim12000$) and the cube making takes much longer time than the spectrum making.
As a result, the total reduction time of a map with a scan pattern was $\sim$7~minutes with 9 iterations and 20 time-chunks.

We note that some digital spectrometers have already equipped $D>10^{4}$ channels (e.g., XFFTS~\citep{Klein2012}; PolariS~\citep{Mizuno2014}), where a correlated component removal becomes time-consuming.
In such cases, EMPCA would also be a good solution to reduce the computation cost.
According to \citet{Bailey2012}, computation cost of EMPCA is $\mc{O}(MNK^3+MNKD)$, where $M$ is the number of iterations within an EM algorithm (not iterations of a data reduction procedure).
With the number of channels of XFFTS ($D=2^{15}$) and a typical iteration time of $M=10^{2}$, the computation cost of EMPCA is much smaller than that of PCA by about two orders of magnitude.
The reduction of the computation cost by EMPCA will be demonstrated in the next study.

\section{Conclusions}
\label{s:conclusions}

In this paper, we propose a new observing method for millimeter and submillimeter spectroscopy to achieve high observation efficiency ($\eta\msb{obs}>0.9$) and baseline stability based on the correlated noise removal technique.
FMLO, our proposed method, employs spectral correlation of sky emission for instantaneous removal of the emission on a timestream, while astronomical signals are frequency-modulated by an LO whose frequency is fast-sweeped ($\sim$10~Hz).
The conclusions are as follows:

\begin{itemize}
  \item
  We show that the correlated noise removal technique used in continuum imaging and CMB experiments can be applied to (sub-)millimeter spectroscopy by frequency modulation of a spectral band, which is realized by an FMLO.
  \item
  We establish the principle of the FMLO method by introducing a mathematical expression of a timestream and its modulation and demodulation.
  As a specific advantage of the FMLO method, we also express the software-based sideband separation as a result of reverse-demodulation of a timestream.
  \item
  We develop an FMLO observing system and install it on the TZ front-end receiver of the Nobeyama 45-m telescope.
  We achieve accurate time synchronization between the telescope' s 1~pps clock and frequency modulation of a digital signal generator that generates the first LO signal of the TZ.
  \item
  We develop a software-based data reduction procedure for the FMLO method, which employs PCA for the correlated noise removal and an iterative algorithm for accurate estimation of both correlated components and astronomical signals in a timestream.
  \item
  We conduct single-pointed and mapping observations of Galactic sources with both the FMLO and PSW methods.
 We demonstrate that the observation efficiency of the FMLO method is dramatically higher than that of the PSW method:
  It is at least 3.0 times and 1.2 times better in single-pointed observations and mapping observations, respectively, while the obtained intensities of spectra or maps are consistent between the two methods.
  The efficiency of the mapping observation could be improved in a larger ($\sim$deg$^{2}$) scale mapping, based on our calculation of mapping design.
  \item
  We find that the estimated correlated components contribute noise to a cleaned timestream, although it is a small contribution compared with that of an off-point measurement.
  We also have to consider the effect of atmospheric line emission in some observing frequencies.
  It will be necessary to introduce the weighted PCA method so that correlated components can be smoothed or the atmospheric line emission deweighted, which will be further investigated in the next study.
\end{itemize}

\begin{ack}
  We thank the anonymous referee for fruitful comments.
  This work is supported by KAKENHI (Nos. 23840007, 25103503, 15H02073, 17H06130, and 15J05101).
  This paper makes use of data taken by the Nobeyama 45-m radio telescope.
  The Nobeyama 45-m radio telescope is operated by the Nobeyama Radio Observatory, a branch of the National Astronomical Observatory of Japan.
  We made use of SciPy\footnote{Jones et al. 2001--, Scipy: Open Source Scientific Tools for Python, \url{https://www.scipy.org}}, NumPy~\citep{Vanderwalt2011}, xarray~\citep{Hoyer2017}, Numba~\citep{Lam2015}, scikit-learn~\citep{Pedregosa2011}, Astropy~(Astropy Collaboration et al. \yearcite{Astropy2013}, \yearcite{Astropy2018}), and matplotlib~\citep{Hunter2007} for the development of \texttt{FMFlow}.
  We made use of CASA~\citep{McMullin2007} for making intengrated intensity maps and spectra of both OTF and FMLO mapping observations.
\end{ack}

\end{document}